\documentclass[aps,prb,superscriptaddress,amsfonts,showpacs,twocolumn,eqsecnum]{revtex4}
\usepackage{epsfig, graphicx,psfrag,amsmath,amssymb,float}
\newcommand{\be}{\begin{equation}}
\newcommand{\ee}{\end{equation}}
\newcommand{\bea}{\begin{eqnarray}}
\newcommand{\eea}{\end{eqnarray}}
\newcommand{\ket}{\rangle}
\newcommand{\bra}{\langle}

\begin{document}

\title{Spectral function of spinless fermions on a one-dimensional lattice}
\author{Rodrigo G. Pereira}
 \affiliation{Kavli Institute for Theoretical Physics, University of California, Santa Barbara, CA 93106, USA}
\author{Steven R. White}
\affiliation{Department of Physics and Astronomy, University of California,
 Irvine, CA 92697, USA}
\author{Ian Affleck}
\affiliation{Department of Physics and Astronomy, University of British
  Columbia, Vancouver, BC, Canada V6T 1Z1}
\date{\today}
\begin{abstract}
We study the spectral function of interacting one-dimensional fermions for an integrable lattice model  away from half-filling.  The divergent power-law singularity of the spectral function near the single-particle or single-hole energy is described by an effective x-ray edge type model. At low densities and for momentum near the zone boundary, we find a second divergent singularity at higher energies which is associated with a two-particle bound state. We use the Bethe ansatz solution of the model to calculate the exact singularity exponents   for any momentum and for arbitrary values of chemical potential and interaction strength in the critical regime. We relate the singularities of the spectral function to the long-time decay of the fermion Green's function and compare our predictions with numerical results from the time-dependent  density matrix renormalization group (tDMRG). Our results show that the tDMRG method is able to provide accurate time decay exponents in the cases of power-law decay of the Green's function. Some implications for the line shape of the dynamical structure factor away from half-filling are also discussed. In addition, the spectral weight of the Luttinger liquid result for the dynamical structure factor of the Heisenberg model at zero field is compared with the exact two-spinon contribution. 
\end{abstract}
\pacs{75.10.Pq, 71.10.Pm}
\maketitle
\section{Introduction}
 
The single-particle spectral function provides direct information about the elementary excitations of a system.  A recent experiment\cite{stewart} has demonstrated the possibility of measuring the spectral function of strongly interacting cold atomic gases using momentum-resolved radio-frequency spectroscopy.\cite{dao}  This technique can be applied to various highly controllable realizations of condensed matter systems,  motivating theoretical predictions for the line shapes which can be measured in such experiments.

An example of interaction-induced exotic behavior reflected in the single-particle spectral function is that of one-dimensional (1D) fermionic systems. 1D metals  distinguish themselves from higher-dimensional Fermi liquids by the absence of stable quasiparticles in the low-energy regime,\cite{giamarchi} in which they are generally described as Luttinger liquids.\cite{haldane} While in a Fermi liquid the main effect of weak repulsive interactions on the spectral function is to broaden  the quasiparticle peak,\cite{nozieres} in a Luttinger liquid the  peak is replaced by an asymmetric  power-law singularity above the single-particle energy, with an exponent which depends on the interaction strength.\cite{meden,voit}  The case of spin-1/2 fermions, for which two spin-charge separated singularities are predicted, is relevant for quasi-1D conductors.\cite{bourbonnais}  The simpler  case of spinless fermions, which shall be the subject of this paper, can be realized in fully spin-polarized ultracold fermionic gases.\cite{gunter}

The Luttinger liquid result for the spectral function is only asymptotically exact in the low energy limit because it relies on the approximation of linear dispersion of the  particle and hole excitations about the Fermi points. It is known that this approximation yields the correct long distance asymptotic behavior of correlation functions because band curvature terms in the Hamiltonian are formally irrelevant.\cite{haldane} However,  recently it has been pointed out that dispersion nonlinearity can modify the line shape of dynamical correlation functions in the vicinity of the single-particle energy, $\omega=\epsilon(k)$.\cite{pustilnik,khodas} In fact, in the case of weakly interacting spinless fermions with parabolic dispersion, the support of the spectral function $A(k,\omega)$ extends below the single-particle energy. Khodas et al.\cite{khodas} have shown that, for generic two-body  interactions,   the coupling of the single particle to a continuum  of excitations with multiple particle-hole pairs  gives rise to a decay rate and rounds off the  singularity on the single-particle energy. In one dimension, the decay rate $1/\tau$ is remarkably small because it depends on three-body scattering processes. For momentum $k$ near and above the Fermi momentum $k_F$, it vanishes as $1/\tau\sim V_0^2(V_0-V_{k-k_F})^2(k-k_F)^4/m^3v_F^6$, where  $V_q$ is the Fourier transform of the interaction potential, $v_F$ is the Fermi velocity and $m$ is the effective mass. For $k\approx k_F$, the other energy scale set by band curvature is $\delta\omega=(k-k_F)^2/m$. For $|\omega-\epsilon(k)|\sim 1/\tau$, the particle spectral function resembles the Lorentzian shape expected for a Fermi liquid with quasiparticle decay rate $1/\tau$. For $1/\tau\ll |\omega-\epsilon(k)|\ll\delta\omega$, the spectral function assumes the form of a two-sided power-law singularity with a different exponent than the one predicted by Luttinger liquid theory. The nature of this power law in the vicinity of the single-particle energy is understood by analogy with the problem of the x-ray edge singularity in metals.\cite{pustilnik} The power law with the Luttinger liquid exponent is only recovered for $ \delta\omega\ll\omega-\epsilon(k)\ll k_F^2/m$. The crossover between the two power laws is described by a universal scaling function of $[\omega-\epsilon(k)]/\delta\omega$.\cite{imambekov}

Adding a quadratic term to the dispersion also breaks particle-hole symmetry. In general, the hole  contribution to the spectral function becomes qualitatively different from the particle contribution once band curvature effects are taken into account. In the continuum model, the single hole  with momentum $-k_F<k<k_F$ is stable since it coincides with the lower threshold of the multiparticle continuum. As a result, there is an exact  power-law singularity at the energy of the single hole. However, the exponent in the range $|\omega-\epsilon(k)|\ll\delta\omega$ is again not the one predicted by Luttinger liquid theory.\cite{khodas}

The study of the singularities of spectral functions using x-ray edge type effective models is not limited to  low energies or to weak interactions. In fact, in the case of integrable models it is even possible to compute  exact exponents and energy thresholds for arbitrary momentum and interaction strength.  The key is to extract these parameters from the exact finite size spectrum calculated by Bethe ansatz (BA). \cite{woynarovich,korepin} The phase shifts that appear in the finite size spectrum fix the parameters of the effective field theory, which can then be used to calculate correlation functions. This was done for the edge singularities of the dynamical structure factor of the spin-1/2 XXZ chain,\cite{pereira} or equivalently for spinless fermions on a lattice, \cite{cheianov} and for both the dynamical structure factor and the spectral function of interacting bosons. \cite{imamb2} The idea of extracting exponents of finite-energy spectral functions from the BA had appeared earlier in the pseudofermion dynamical theory for the 1D Hubbard model.\cite{carmelo} The field theory prediction for the singular features of the spectral function can be combined with numerical methods such as the time-dependent density matrix renormalization group (tDMRG)\cite{tDMRG} to produce high resolution line shapes.\cite{pereira} 

Numerical results for the spectral function of the integrable model of spinless fermions on a lattice should thus offer a quantitative test for the predictions of effective x-ray edge models, in particular for the line shape proposed by Khodas et al.\cite{khodas} Two remarks are in order. The first one is that  integrable models are \emph{non-generic} in the sense that they possess an infinite number of local conserved quantities.\cite{sutherland} What makes these models amenable to the BA is precisely that all scattering processes can be factorized into a series of two-body collisions. It is not clear whether integrable models can have a nonzero decay rate $1/\tau$ which smoothes out the singularities of the spectral function. In principle, the question of a small versus vanishing decay rate due to interactions can  be addressed experimentally, since cold atom systems have been shown to exhibit very low dissipation and to be approximately described by integrable models.\cite{kinoshita}

The second difference from the continuum model is that placing the particles in a one-dimensional lattice introduces effects which are not observed in free space. A noteworthy example is the appearance of stable repulsively bound states, which has been demonstrated experimentally in the case of bosons.\cite{winkler} Bound  molecules of spin-polarized fermions have been created artificially by sweeping across a $p$-wave Feshbach resonance.\cite{gaebler}  ``Antibound" $p$-wave molecules in the case of repulsive interactions should arise naturally as excited states in a lattice. An important question is whether these bound states can produce a strong response in the \emph{single-particle} spectral function.

The purpose of this work is to study the spectral function for the integrable lattice model of 1D spinless fermions with repulsive nearest-neighbor interaction. We are particularly interested in the regime of high energies and strong interactions, where lattice effects are most important. 

The paper is organized as follows. In Sec. \ref{sec:kin}, we discuss the kinematics of the model with a cosine dispersion. In Sec. \ref{sec:ham}, we  present the effective x-ray edge Hamiltonian which describes the behavior of the  spectral function near the single-particle or single-hole energy.  In Sec. \ref{sec:bound}, we show that repulsively bound states can give rise to additional divergent singularities at high energies, which   are more pronounced at low densities and for momentum near the zone boundary. In Sec. \ref{sec:integrable}, we address the effects of integrability on the spectral function and on the  long-time decay of the particle Green's function. We show that the broadening of the singularity at the single-particle energy in the regime considered by Khodas et al. can be recovered by adding an irrelevant interaction term to the effective Hamiltonian. In addition, we  argue that in the integrable lattice model  a single particle with low velocity can have a nonzero decay rate due to two-body scattering processes. In Sec. \ref{sec:BA}, we derive the formulas for the exact singularity exponents using the BA solution.  In Sec. \ref{sec:DMRG}, we compare  the analytical predictions for the long-time decay  of the fermion Green's function with high precision data from the tDMRG method.  In Sec. \ref{sec:DSF}, we discuss the implications of our results  for the dynamical structure factor away from half-filling. Some concluding remarks are offered in Sec. \ref{sec:discuss}. There are, in addition, three appendices. Appendix A contains the detailed derivation of the finite size spectrum from BA. Appendix B proves the equivalence of our formulas for the singularity exponents to those of Ref. \onlinecite{cheianov}. Finally, in Appendix C we compare the Luttinger liquid result for the dynamical structure factor near $q=\pi$ to the result obtained in the two-spinon approximation, correcting the prefactor found in Ref. \onlinecite{Karbach}.

\section{Kinematics of the lattice model\label{sec:kin}}
We consider the model of spinless fermions with nearest-neighbor interaction
\begin{widetext}
\be
H-\mu N=\sum_{j=1}^{L}\left[-(\psi^\dagger_j\psi^{\phantom\dagger}_{j+1}+\psi^\dagger_{j+1}\psi^{\phantom\dagger}_{j})+V\left(n_j-\frac{1}{2}\right)\left(n_{j+1}-\frac{1}{2}\right)-\mu n_j\right].\label{model}
\ee
\end{widetext}
Here, $\psi_j$ is the operator that annihilates a fermion at site $j$,  $L$ is the number of sites, taken to be even, $V>0$ is the strength of the repulsive interaction, $n_j=\psi^\dagger_j\psi^{\phantom\dagger}_{j}$ is the number operator at site $j$ and $\mu$ is the chemical potential. The lowest energy state for $N$ fermions is unique if we impose periodic (antiperiodic) boundary conditions for $N$ odd (even).  At half-filling, $n=\bra n_j\ket  =1/2$, the Hamiltonian is invariant under the particle-hole transformation $\psi^{\phantom\dagger}_j\to(-1)^j \psi_j^\dagger$. 

The Hamiltonian of Eq. (1) is equivalent to the anisotropic (XXZ) spin-1/2 chain and is BA integrable.\cite{orbach} The ground state phase diagram as a function of $V$ and $\mu$ is known.\cite{sutherland,takahashi} In this paper we will concern ourselves with the region $0\leq V<2$ and $|\mu|<2+V$, where the system is in a gapless phase with power-law decaying  equal-time correlation functions and  low-energy physics described by the Luttinger model.\cite{giamarchi} 

The fermion spectral function is defined as
\be
A(k,\omega)=-\frac{1}{\pi}\,\textrm{Im}\,G_{ret}(k,\omega),
\ee
where \be
G_{ret}(k,\omega)=-i\int_0^\infty dt\,e^{i(\omega+i\eta)t}\bra\{\psi^{\phantom\dagger}_k(t),\psi^\dagger_k(0)\}\ket
\ee
is the retarded single-particle Green's function.\cite{mahan} Here $\psi_k=\sum_je^{-ikj}\psi_j$ annihilates a fermion with momentum $k$. The spectral function contains both particle and hole contributions\be
A(k,\omega)=A_p(k,\omega)+A_h(k,\omega),
\ee
which can be written in the form \bea
A_p(k,\omega)&=&\sum_\alpha \left|\bra\alpha|\psi_k^\dagger|0\ket\right|^2\delta(\omega-E_\alpha+E_0),\label{Apart}\\
A_h(k,\omega)&=&\sum_\alpha \left|\bra\alpha|\psi_k^{\phantom\dagger}|0\ket\right|^2\delta(\omega+E_\alpha-E_0),\label{Ahole}
\eea
where  $E_0$ is the ground state energy and $\left|\alpha\right\ket$  is an exact eigenstate of the Hamiltonian (\ref{model}) with energy $E_\alpha$. 
 
In the noninteracting, $V=0$ case, the Hamiltonian (\ref{model}) reduces to the tight-binding model:\be
H_0-\mu N=\sum_k\epsilon^{(0)}(k)\psi^\dagger_k\psi^{\phantom\dagger}_k,\label{Hnonint}
\ee
where $\epsilon^{(0)}(k)=-2\cos(k)-\mu$ is the free fermion dispersion and $k\in(-\pi,\pi]$. The ground state is constructed by filling up the single-particle states up to the Fermi level, $\epsilon^{(0)}(k_F)=0$. In terms of fermion density, the Fermi momentum reads $k_F=\pi n$. Only excited states with a single particle with momentum $k$ above the Fermi level couple to the ground state via the operator $\psi^\dagger_k$ and contribute to $A_p(k,\omega)$. As a result, for $V=0$:\be
A^{(0)}_p(k,\omega)=\theta(|k|-k_F)\delta(\omega-\epsilon^{(0)}(k)).
\ee
Likewise, for the hole contribution,\be
A^{(0)}_h(k,\omega)=\theta(k_F-|k|)\delta(\omega-\epsilon^{(0)}(k)).
\ee
Therefore, in the noninteracting case, the spectral function is given by a delta function peak at the energy of the single particle (for $k_F<|k|<\pi$) or single hole (for $|k|<k_F$).

In the interacting case, there are nonzero matrix elements between the ground state and excited states with multiple particle-hole pairs, which then contribute to the spectral function. Hereafter we consider the case $k_F<\pi/2$ ($\mu<0$). (From this the case $k_F>\pi/2$ can be understood by exchanging particles and holes.) Since $A(k,\omega)=A(-k,\omega)$, we also assume $0<k<\pi$. Although the renormalized dispersion may deviate  from the cosine form (unlike the case of parabolic dispersion protected by Galilean invariance; {\it c.f.} Ref. \onlinecite{khodas}), in this section we assume a  cosine dispersion in order to discuss the support of $A(k,\omega)$. This is approximately valid in the limit of weak interaction. (However, the discussion does not depend qualitatively on this assumption; the important feature is simply the existence of a single inflection point above the Fermi surface. The exact dispersion can be calculated from the BA, as will be done in Sec. \ref{sec:DMRG}.)

Let us first focus on the particle contribution $A_p(k,\omega)$. For nonzero interactions,  $A_p(k,\omega)$ is always nonzero below the energy of the single-particle excitation. Below half-filling, the positive curvature of the dispersion below the Fermi points implies that the minimum energy for fixed total momentum $k$ must correspond to a state with $|r|+1$ particles at the Fermi point with momentum $\pm k_F$, $|r|-1$ holes at the Fermi point with momentum $\mp k_F$, and one deep hole with momentum $-k_F<k^{(r)}_h<k_F$ such that\be
 (2rk_F-k_h^{(r)})\textrm{ mod}\, 2\pi =k.\label{deepholethresh}
\ee

\begin{figure}
\includegraphics*[width=\columnwidth,scale=1.0]{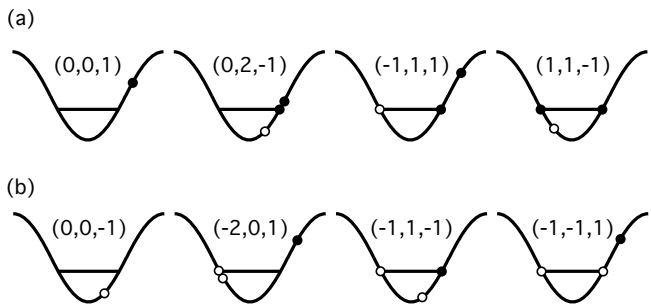}

\caption{Some excitations which define the edges of the support of the spectral function: a) excitations with charge +1, which contribute to $A_p(k,\omega)$; b) excitations with charge $-1$, which contribute to $A_h(k,\omega)$ (see text for notation).\label{fig:excit}}
\end{figure}

Excitations with a single high energy particle or hole can be labeled by three numbers $(N_L,N_R,N_d)$. $N_L$ ($N_R$) is the number of particles created near the left (right) Fermi point; $N_{L,R}>0$ denotes particles and $N_{L,R}<0$ denotes holes. $N_d=1$ in the case of a high energy particle and $N_d=-1$ in the case of a high energy hole. The excitations that contribute to $A_p(k,\omega)$ have charge +1, i.e. $N_L+N_R+N_d=1$. For instance, the excitations that define the deep hole thresholds in Eq. (\ref{deepholethresh}) are labeled $(-r+1,r+1,-1)$ (see Fig. \ref{fig:excit}).

\begin{figure}
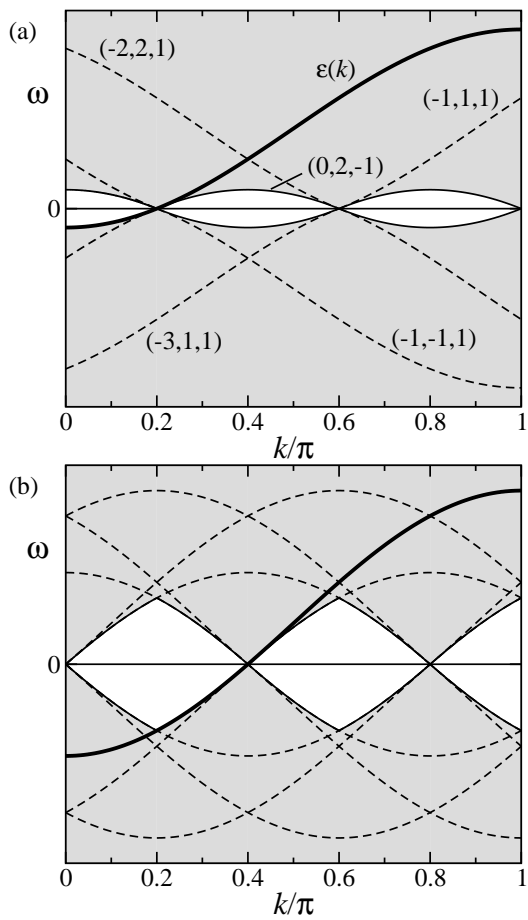

\includegraphics*[width=0.8\hsize,scale=1.0]{fig2a.eps}
\includegraphics*[width=0.8\hsize,scale=1.0]{fig2b.eps}

\caption{Support of the spectral function $A(k,\omega)$ for weakly interacting fermions with a cosine dispersion (the support is the shaded area), for two values of Fermi momentum: a) $k_F=\pi/5$; b) $k_F=2\pi/5$. The positive frequency part corresponds to the particle contribution $A_p(k,\omega)$ and  the negative frequency part to the hole contribution $A_h(k,\omega)$. The lines drawn are the dispersion curves for the excitations with multiple particles and holes at the Fermi points and a single high-energy particle or hole,  labeled as in Fig. \ref{fig:excit} (see also main text). The thicker solid line indicates the dispersion of the single particle or single hole, $\omega=\epsilon(k)$. The other solid lines indicate the  edges of the support, where $A(k,\omega)$ vanishes.  \label{fig:support}}
\end{figure}

Fig. \ref{fig:support} shows the support of the spectral function for two rational values of $k_F/\pi$, $k_F=\pi/5$ and $k_F/\pi=2/5$.  Since in the  lattice  momentum is only defined mod $2\pi$,  for every $k$ value there is an infinite number of deep hole type excitations, corresponding to different choices of $r$, such that Eq. (\ref{deepholethresh}) is verified. Nonetheless, for $k_F/\pi$ rational, for general $k$ there is  a finite lower threshold for the support of $A_p(k,\omega)$. The reason is that, for $k_F/\pi=p/q$ ($p,q$ integers), the energy of the deep hole thresholds, \bea
\omega_{(r+1,-r+1,-1)}(k)&=&-\epsilon^{(0)}(k_h^{(r)})\nonumber\\&=&2\cos\left(2\pi\frac{rp}{q}-k\right)+\mu,\eea  is periodic in $r$ with period $q$.

\begin{figure}
\includegraphics*[width=0.8\hsize,scale=1.0]{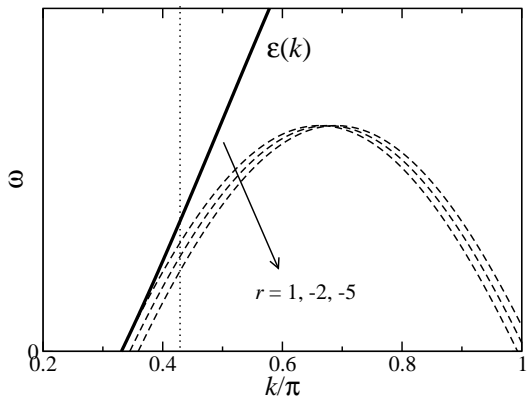}
\caption{Thresholds for deep hole type excitations for $k_F/\pi=1/3-\delta$, $\delta\ll1$. The solid line represents the single-particle dispersion. For irrational values of $k_F/\pi$ and any given $k$, one can construct a series  of ``thresholds" $(-r+1,r+1,-1)$, with increasing number of low-energy particles and holes, $2|r|$, such that the energy  approaches zero. For $k=\pi/2$, we have $\omega_{(0,2,-1)}>\omega_{(3,-1,-1)}>\omega_{(6,-4,-1)}>...$ .  The lower edge vanishes for all $k$ and the support of $A(k,\omega)$ covers the entire $(k,\omega)$ plane.\label{fig:irrational}}
\end{figure}

If $k_F/\pi$ is irrational, the energy of the deep hole thresholds is not periodic in $r$. This is equivalent to the problem of irrational rotations on the unit circle.\cite{nadkarni} In this case, for any $k$ there are infinitely many nondegenerate thresholds $(-r+1,r+1,-1)$ and the energy can be made arbitrarily low by taking $|r|$ sufficiently large. In other words, for  $k_F/\pi$ irrational,  we can move the momentum of the deep hole  arbitrarily close to the Fermi surface (at  $\pm k_F$) by shifting it by multiples of $2k_F$ (considering only the cases where the shift takes the hole to an allowed region, below the Fermi surface). Fig. \ref{fig:irrational} illustrates the case $k_F/\pi=1/3-\delta$ irrational with $\delta\ll1$. Consider, for example, $k=\pi/2$. Since $k_F$ is not commensurate with $\pi$, the energy of the $(3,-1,-1)$ threshold falls slightly below the energy of the $(0,2,-1)$ threshold. Continuing with the series, we find that the $(6,-4,-1)$ threshold  falls at an even lower energy than $(3,-1,-1)$. This series of thresholds with decreasing energy implies that the support of $A_p(k,\omega)$ extends down to zero energy for all $k$ if $k_F/\pi$ is irrational. 

The edges of the support of the hole contribution $A_h(k,\omega)$ can be discussed in a similar fashion. Some  excitations with  charge $N_R+N_L+N_d=-1$ are illustrated in Fig. \ref{fig:excit}b. The notation is such that the dispersion relation for the $(N_L,N_R,N_d)$ excitation as a function of the momentum of the high energy particle or hole is the continuation of the dispersion of the $(-N_R,-N_L,-N_d)$ excitation to negative energies. The upper thresholds of $A_h(k,\omega)$ (in the sense of a maximum negative frequencies above which $A_h(k,\omega)$ is zero) are defined by deep hole excitations labeled $(-r,r,-1)$. The support of $A_h(k,\omega)$ for two cases of rational $k_F/\pi$  is  shown in Fig. \ref{fig:support}. Note that, unlike the case of parabolic dispersion,\cite{khodas} the single hole is not always at the edge of a continuum (see, for example, the threshold for $k_F=2\pi/5$ and $k<\pi/5$).

\section{Effective Hamiltonian for  singularity near the single-particle energy\label{sec:ham}}
The behavior of the spectral function near the energies of the single particle or single hole and near the edges of the support for  nonlinear dispersion can be studied using the methods developed in Refs. \onlinecite{khodas,pustilnik}. The procedure consists of integrating out high-energy modes of the fermionic field and introducing subbands which include not only the standard low energy modes at $\pm k_F$, but also  the modes around the momentum of the high energy particle (or hole) whose energy defines the threshold. This leads to an effective model in which the high energy particle acts as a mobile impurity coupled to the Luttinger liquid modes.\cite{castroneto,kawakami,balents} The anomalous exponents at the thresholds are associated with the phase shifts at the Fermi surface due to the creation of the high energy particle, in analogy with the x-ray edge singularity in metals.\cite{mahanxray,ND,schotte}

Here we focus  on the power-law singularity around the single-particle or single-hole energy $\omega=\epsilon(k)$. We take the continuum limit of the Hamiltonian in Eq. (\ref{model}) and introduce the fermionic field $\psi(x)$.  In order to study the singularity at the energy of the single particle with momentum $k$, we expand \be
\psi(x)\sim e^{ik_Fx}\psi_R(x)+e^{-ik_Fx}\psi_L(x)+e^{ikx}d(x),\label{modeexp}
\ee
where $\psi_{R,L}(x)$ are right and left movers with momentum near $\pm k_F$ and $d(x)$ is the high energy mode defined from  states with momentum near $k$, for $k_F<k<\pi$. Starting from Eq. (\ref{model}), we linearize the free fermion dispersion $\epsilon^{(0)}(k)$ about $\pm k_F$ and $k$.  In the continuum limit, the Hamiltonian density for the hopping term in Eq. (\ref{model}) becomes\be
\mathcal{H}_0=-iv_F(\psi^\dagger_R\partial_x\psi_R^{\phantom\dagger}-\psi^\dagger_L\partial_x\psi_L^{\phantom\dagger})+d^\dagger(\varepsilon_0-iu_{0}\partial_x)d^{\phantom\dagger},
\ee
where $v_F=2\sin k_F$ is the bare Fermi velocity, $\varepsilon_0=\epsilon^{(0)}(k)$ and $u_{0}=2\sin k$ is the bare velocity of the high energy particle. Bosonizing right and left movers in the form \be 
\psi_{R,L}(x)\sim \frac{1}{\sqrt{2\pi\eta}}e^{-i\sqrt{2\pi}\phi_{R,L}(x)},
\ee
where $\phi_{R,L}$ are the right- and left-moving components of a bosonic field and $\eta$ is a short-distance cutoff, we can write\be
\mathcal{H}_0=\frac{v_F}{2}\left[(\partial_x\phi_R)^2+(\partial_x\phi_L)^2\right]+d^\dagger(\varepsilon_0-iu_{0}\partial_x)d^{\phantom\dagger}\label{H0bosonized}
\ee
From Eq. (\ref{modeexp}), we have the expansion of the density operator\begin{widetext}\be
n(x)\sim n+\psi^\dagger_R\psi_R^{\phantom\dagger}+\psi^\dagger_L\psi_L^{\phantom\dagger}+d^\dagger  d^{\phantom\dagger}+[e^{i2k_Fx}\psi^\dagger_L\psi_R^{\phantom\dagger}+e^{i(k-k_F)x}\psi^\dagger_Rd^{\phantom\dagger}+e^{i(k+k_F)x}\psi^\dagger_Ld^{\phantom\dagger}+h.c.].\label{densityop}
\ee
Using Eq. (\ref{densityop}) and being careful about the point splitting of the fermionic fields, we find the Hamiltonian density for the interaction term in Eq. (\ref{model}) \bea
\mathcal{H}_{int}&\sim& \frac{V\sin^2k_F}{\pi}\left(\partial_x\phi_R-\partial_x\phi_L\right)^2-\frac{2V}{\pi}\sin k_F\left(\cos k \,d^\dagger d^{\phantom\dagger}+i\sin k \,d^\dagger\partial_xd^{\phantom\dagger}\right)\nonumber\\
&&-\frac{4V\sin^2[(k-k_F)/2]}{\sqrt{2\pi}}\partial_x\phi_R d^\dagger d^{\phantom\dagger}+\frac{4V\sin^2[(k+k_F)/2]}{\sqrt{2\pi}}\partial_x\phi_L d^\dagger d^{\phantom\dagger},\label{Hintbosonized}
\eea
where  we have set $\eta=1$ (following Ref. \onlinecite{giamarchi}) and omitted irrelevant interaction terms and the renormalization of the chemical potential. Note that the coupling to the $d$ modes explicitly breaks the parity symmetry of the original Hamiltonian.
\end{widetext}

Combining Eqs. (\ref{H0bosonized}) and (\ref{Hintbosonized}), we obtain\bea
\mathcal{H}&=&\frac{vK}{2}\left(\partial_x\tilde\theta\right)^2+\frac{v}{2K}\left(\partial_x\tilde\phi\right)^2+d^\dagger \left(\varepsilon-iu\partial_x\right)d^{\phantom\dagger}\nonumber\\
&&+\left(\frac{\tilde\kappa_{L}-\tilde\kappa_{R}}{2\sqrt{\pi}}\partial_x\tilde\theta+\frac{\tilde\kappa_{L}+\tilde\kappa_{R}}{2\sqrt{\pi}}\partial_x\tilde\phi\right)d^\dagger d^{\phantom\dagger},\label{Hbeforescaling}
\eea
where $\tilde\phi$ and $\tilde\theta$ are defined by\be
\phi_{R,L}=\frac{\tilde\theta\mp\tilde\phi}{\sqrt{2}},
\ee
such that $[\tilde\phi(x),\partial_x\tilde\theta(x^\prime)]=i\delta(x-x^\prime)$. To first order in $V$, the parameters in Eq. (\ref{Hbeforescaling}) are\bea
\frac{v}{v_F}&\approx&\frac{u}{u_0}\approx\frac{\varepsilon}{\varepsilon_0}\approx1+\frac{V}{\pi}\sin k_F,\\
K&\approx&1-\frac{V}{\pi}\sin k_F,\\
\tilde\kappa_{R,L}&\approx&2V\sin^2[(k\mp k_F)/2].
\eea
The parameters $v$ and $K$ are the familiar renormalized velocity and Luttinger parameter of the Luttinger model.\cite{giamarchi} The renormalized energy and velocity of the high-energy particle are given by $\varepsilon$ and $u$, respectively. Rescaling the bosonic fields by \be
\tilde\phi=\sqrt{K}\phi\quad,\quad\tilde\theta=\theta/\sqrt{K},
\ee
and introducing the chiral components of the rescaled field\be
\varphi_{R,L}=\frac{\theta\mp\phi}{\sqrt{2}},
\ee
we arrive at the effective Hamiltonian density\bea
\mathcal{H}&=&\frac{v}{2}\left[\left(\partial_x\varphi_R\right)^2+\left(\partial_x\varphi_L\right)^2\right]+d^\dagger \left(\varepsilon-iu\partial_x\right)d^{\phantom\dagger}\nonumber\\
&&+\frac{1}{\sqrt{2\pi K}}\left(\kappa_{L}\partial_x\varphi_L-\kappa_{R}\partial_x\varphi_R\right)d^\dagger d^{\phantom\dagger},\label{Hafterscaling}
\eea
where\be
\kappa_{R,L}=\left(\frac{1+K}{2}\right)\tilde\kappa_{R,L}-\left(\frac{1-K}{2}\right)\tilde\kappa_{L,R}
\ee
are the coupling constants between the $d$ particle and the Fermi surface modes.  

The Hamiltonian of Eq. (\ref{Hafterscaling}) is not restricted to weak interactions. It can be regarded as exact (up to irrelevant operators) if all the parameters are taken to be the renormalized ones, to be fixed by a method that is nonperturbative in $V$. In fact, the exact $v$ and $K$ are fixed by the low-energy spectrum and compressibility computed from the BA solution of Hamiltonian (\ref{model}).\cite{korepin} In Sec. \ref{sec:BA}, we will describe how  $\varepsilon$, $u$ and $\kappa_{R,L}$ can also be fixed with the help of the BA solution, so that we end up with a parameter-free effective theory for the singularities of the spectral function. 

We can decouple the $d$ particle from the bosonic fields by performing a unitary transformation\be
U=\exp\left[-\frac{i}{\sqrt{2\pi K}}\int_{-\infty}^{+\infty}dx\left(\gamma_R\varphi_R+\gamma_L\varphi_L\right)d^\dagger d^{\phantom\dagger}\right],\label{unitary}
\ee
with the parameters\be
\gamma_{R,L}(k)=\frac{\kappa_{R,L}(k)}{v\mp u(k)}.\label{gammau-v}
\ee
For $V\ll1$, we have the weak coupling expressions\be
\gamma_{R,L}\approx\mp\frac{V}{2}\frac{\sin[(k\mp k_F)/2]}{\cos[(k\pm k_F)/2]}.\label{gammaweakcoup}
\ee
In terms of the transformed fields $\bar\varphi_{R,L}=U\varphi_{R,L}U^\dagger$, $\bar{d}=UdU^\dagger$,\bea
\partial_x\varphi_{R,L}&=&\partial_x\bar\varphi_{R,L}\pm\frac{\gamma_{R,L}}{\sqrt{2\pi K}}\bar{d}^\dagger \bar{d}^{\phantom\dagger}\label{phishift},\\
d^{\phantom\dagger}&=&\bar{d}^{\phantom\dagger}\exp\left[-\frac{i}{\sqrt{2\pi K}}\left(\gamma_R\bar\varphi_R+\gamma_L\bar\varphi_L\right)\right],
\eea
the Hamiltonian (\ref{Hafterscaling}) becomes noninteracting (up to irrelevant operators)\be
\mathcal{H}=\frac{v}{2}\left[\left(\partial_x\bar\varphi_R\right)^2+\left(\partial_x\bar\varphi_L\right)^2\right]+\bar{d}^\dagger \left(\varepsilon-iu\partial_x\right)\bar{d}^{\phantom\dagger}.\label{Hdecoupled}
\ee
We are now able to calculate the time-dependent particle Green's function\be
G_p(k,t)=\bra\psi^{\phantom\dagger}_k(t)\psi^\dagger_k(0)\ket,\label{Gpktparticle}
\ee
using the Hamiltonian of Eq. (\ref{Hdecoupled}). 
For $t\gg\varepsilon^{-1}$, we can  use the mode expansion of Eq. (\ref{modeexp}) and write\be
G_p(k,t)\sim\int dx\,\bra d(x,t)d^\dagger(0,0)\ket. 
\ee
In terms of the decoupled fields,\bea
G_p(k,t)&\sim&\int dx\,\bra \bar{d}(x,t)\bar{d}^\dagger(0,0)\ket\nonumber\\&&\times\bra e^{-i\sqrt{2\pi\nu_R}\bar\varphi_R(x,t)}e^{i\sqrt{2\pi\nu_R}\bar\varphi_R(0,0)}\ket\nonumber\\
&&\times\bra e^{-i\sqrt{2\pi\nu_L}\bar\varphi_L(x,t)}e^{i\sqrt{2\pi\nu_L}\bar\varphi_L(0,0)}\ket,
\eea
where \be
\nu_{R,L}(k)=\frac{1}{K}\left(\frac{\gamma_{R,L}(k)}{2\pi}\right)^2\label{nus}
\ee
are the anomalous exponents. The ground state of Hamiltonian (\ref{Hdecoupled}) has $N_d=0$ high-energy particles, and a single $\bar{d}$ particle is created in the transition. The free propagator for the $d$ particle is\be
G_p^{(0)}(x,t)=\bra \bar{d}^{\phantom\dagger}(x,t)\bar{d}^\dagger(0,0)\ket=e^{-i\varepsilon t}\delta(x-ut).\label{propfreepart}
\ee 
The correlation functions of free bosonic fields can be calculated by standard methods.\cite{giamarchi} We find\be
G_p(k,t)\sim e^{-i\varepsilon t}\left[\frac{i}{(u-v)t+i\eta}\right]^{\nu_R}\left[\frac{-i}{(u+v)t-i\eta}\right]^{\nu_L}.\label{Gktpart}
\ee
The long-time asymptotic behavior of the particle Green's function is then given by\be
G_p(k,t)\sim\frac{\exp\left\{-i\varepsilon t-i\frac{\pi}{2}[\nu_L-\textrm{sign}(u-v)\nu_R]\right\}}{t^{\nu}}.\label{Gktdecay}
\ee
where\be
\nu(k)=\nu_R(k)+\nu_L(k).\label{nunuRnuL}
\ee
Taking the Fourier transform of Eq. (\ref{Gktpart}), we obtain the particle contribution to the spectral function\be
A_p(k,\omega)=\int_{-\infty}^{+\infty}dt\,e^{i\omega t}G_p(k,t). 
\ee
The result depends on the sign of $u-v$. For $u>v$,\be
A_p(k,\omega)\sim \frac{\theta(\omega-\varepsilon)\sin(\pi\nu_L)+\theta(\varepsilon-\omega)\sin(\pi\nu_R)}{|\omega-\varepsilon|^{1-\nu}}.\label{Apsingu>v}
\ee
For $u<v$,\be
A_p(k,\omega)\sim\frac{\theta(\omega-\varepsilon)\sin(\pi\nu)}{|\omega-\varepsilon|^{1-\nu}}.\label{Apsingu<v}
\ee
Therefore, the effective model predicts that  $A_p(k,\omega)$ diverges on both sides of the single-particle energy $\omega=\varepsilon$ if $u>v$, but only from above if $u<v$. In the continuum model,\cite{khodas} the velocity increases monotonically with $k$, thus one always has $u>v$. In the lattice model, the velocity vanishes as $k$ approaches the zone boundaries. In the noninteracting limit, we have $u<v$ for $\pi-k_F<k<\pi$. Note that the result in Eq. (\ref{Apsingu<v}) does not imply that $A_p(k,\omega)$ vanishes below the single-particle energy if $u<v$. As we discussed in Sec. \ref{sec:kin}, the support of the spectral function always extends below the single-particle energy for $k_F<\pi/2$. However, a singular behavior below the single-particle energy only appears if the single-particle excitation is unstable (i.e. the energy is lowered) against the emission of a low-energy particle-hole pair at the Fermi surface.\cite{pustilnik} This is the case of a ``fast particle", with $u>v$.  For a ``slow particle" with $u<v$, decay processes with a \emph{finite} momentum transfer can still lower the energy of the single-particle excitation. Both finite momentum transfer and three-body collision type  processes can smooth out the singularity on the single-particle energy at the scale of a decay rate $1/\tau$.  We shall return to this point in Sec. \ref{sec:integrable}.

Similar results can be derived for the hole Green's function \be
G_h(k,t)=\bra\psi_k^{\dagger}(t)\psi_k^{\phantom\dagger}(0)\ket\label{Ghktdecay}
\ee
by employing a Hamiltonian analogous to Eq. (\ref{Hafterscaling}) and interpreting  $d$  as the operator that annihilates a hole with momentum $k<k_F$. We find\be
G_h(k,t)=\frac{\exp\left\{-i\varepsilon t-i\frac{\pi}{2}[\nu_L+\textrm{sign}(v-u)\nu_R]\right\}}{t^{\nu}},
\ee
where $\varepsilon>0$ and $u>0$ are the renormalized energy and velocity of the hole, and $\nu_{R,L}$ are the corresponding exponents as given by Eq. (\ref{nus}). Typically, we expect $u<v$ for a hole below $k_F<\pi/2$, but it is possible that the effective mass around the Fermi points changes sign in the strongly interacting regime.\cite{pereiraJSTAT} For $u<v$, we  find\be
A_h(k,-\omega)\sim\frac{\theta(\omega-\varepsilon)\sin[\pi(\nu_R+\nu_L)]}{|\omega-\varepsilon|^{1-\nu}}.\label{Ahsingu<v}
\ee

The singularity exponents at the edges of the support defined by excitations with multiple particles and holes (e.g. $(2,0,-1)$ in Fig. \ref{fig:support}) can also be calculated using x-ray edge type effective models as done in Refs. \onlinecite{khodas, imambekov}. In the continuum model, these singularities are always convergent due to phase space constraints. In this paper we are only  interested  in divergent singularities, which are the most apparent feature of the spectral function and govern the long-time behavior of the fermion Green's function. Divergent singularities are also the easiest to investigate with numerical methods.  As we shall discuss in Sec. \ref{sec:bound}, the lattice model has thresholds defined by excitations with more than one high energy particle. When two of these particles form a bound state, additional divergent singularities can arise.

\section{Divergent singularities from bound states\label{sec:bound}}
The non-monotonic momentum dependence  of the velocity  in the lattice model affects the nature of the upper thresholds of multiparticle continua. When an upper threshold is defined by excitations which propagate with the same velocity, the interaction between the high-energy particles and holes can lead to the formation of bound states and strongly affect  the line shape of the spectral function.\cite{pereira} In this section, we examine the possibility that a two-particle one-hole state can give rise to a divergent power-law singularity in the single-particle spectral function when two particles with negative mass bind for $V>0$.

\begin{figure}
\includegraphics*[width=0.95\hsize,scale=1.0]{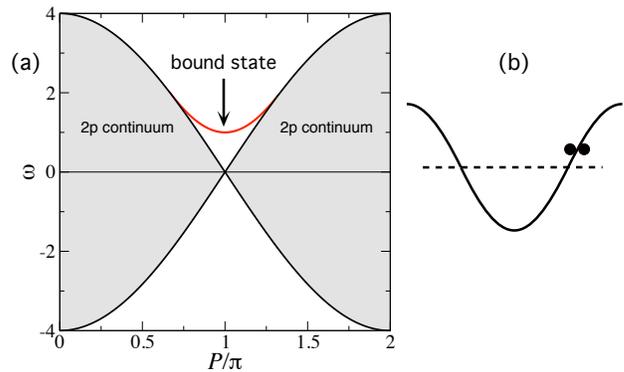}
\caption{(Color online) a) Two-particle continuum and bound state band for the two-body problem on the lattice. b) The upper threshold of the two-particle continuum is defined by two particles with approximately the same momentum and negative effective mass. \label{fig:boundstate}}
\end{figure}

The existence of bound states above the two-particle continuum can be demonstrated directly by solving the two-body problem in the lattice. This is a trivial limit of the BA equations for the eigenstates of Eq. (\ref{model}) for $N=2$ particles.\cite{sutherland} Consider the spinless fermion model (dropping the chemical potential terms)
\be H=-\sum_j (\psi^\dagger_j\psi^{\phantom{\dagger}}_{j+1}+h.c.) +V\sum_j n_j n_{j+1}.\ee 
Consider two-particle eigenstates
\be |\Psi\ket=\sum_{i,j}\phi_{i,j}\psi^\dagger_i\psi^\dagger_j|0\ket,\ee
where we may assume $\phi_{i,j}=-\phi_{j,i}$. The wave function must obey the lattice Schr\"odinger equation 
\bea -\phi_{i+1,j}-\phi_{i-1,j}-\phi_{i,j+1}-\phi_{i,j-1}&=&\mathcal{E} \phi_{i,j},\ \  (|i-j|>1)\nonumber \\
\phi_{i+2,i}-\phi_{i+1,i-1}&=&(\mathcal{E} -V)\phi_{i+1,i},\eea
where $\mathcal{E}$ is the energy eigenvalue. We may separate the center-of-mass and relative coordinates
\be \phi_{j,l}=e^{iP(j+l)/2}f(j-l).\ee
Note that the momentum of this state is $P$. The Schr\"odinger equation becomes
\bea f(r+1)+f(r-1)&=&-{\mathcal{E} \over 2\cos (P/2)}f(r)\nonumber \\
f(2)&=&-{\mathcal{E} -V\over 2\cos (P/2)}f(1).\label{SE}\eea
The most general solution of the first equation is (for $r>0$)
\be f(r)=e^{-\mathcal{Q}r},\ee
where $\mathcal{Q}$ must be either pure real or pure imaginary in order for $\mathcal{E}$ to be real. 
The case of real positive $\mathcal{Q}$ corresponds to a bound state. $\mathcal{Q}$ is given by
\be \cosh \mathcal{Q} = -{\mathcal{E} \over 4\cos (P/2)}.\ee
The second of Eqs. (\ref{SE}) determines $\mathcal{Q}$ and $\mathcal{E}$
\be e^{\mathcal{Q}}=-{V\over 2\cos (P/2)}.\ee
Thus we see that, for $V>0$, a bound state only exists for a range 
of wave vector $P$ where $\cos (P/2)<0$. Requiring $\mathcal{Q}>0$, we see that
\be -\cos (P/2) <V/2.\ee
For small positive $V$, $P/2=\pm (\pi /2+\delta )$ with 
small positive $\delta$. Thus the momentum of this state is 
$P\approx \pi \pm 2\delta$.  The allowed 
range of momentum is
\be \sin \delta <V/2.\label{deltaboundK}\ee
The dispersion relation of this two-particle bound state is
\be \mathcal{E} (P)=V+\frac{2}{V}+\frac{2}{V}\cos P,\ee
with the minimum, $\mathcal{E} =V$, at momentum $P=\pi$. Note that the effective hopping amplitude  is $t_{eff}=2/V>1$, therefore the repulsively bound state is lighter than the free particles. The two-particle continuum in the range of Eq. (\ref{deltaboundK}) extends  over\be
-4\sin\delta<\mathcal{E}<4\sin\delta.
\ee
Since $4\sin\delta<V+(4/V)\sin^2\delta$, the energy of the bound state is above the continuum. In this sense, it is an antibound state (see Fig. \ref{fig:boundstate}a).

The BA solution shows that two-particle bound states also exist in the case of finite fermion density.\cite{sutherland} We are interested in excitations in which a two-particle bound state is added to the ground state of the system with $N$ particles. The momentum range and dispersion relation for this type of excitation can be calculated exactly in the thermodynamic limit, as we will discuss in Sec. \ref{sec:BA}. In the remainder of this section, we work out the effective field theory for the singularities involving two-particle bound states.

We first look more closely at the nature of the upper threshold of the two-particle continuum discussed above. In general,  one could define two subbands with momenta around $k_1$ and $k_2$, with $k_1+k_2=P$ mod $2\pi$, and expand the dispersion relations within each subband in the form\be
\epsilon_{i}(\delta k_i)\approx \varepsilon_{i}+u_{i}\delta k_i+\frac{(\delta k_i)^2}{2m_i},
\ee
for $i=1,2$. The energy of a particle pair with total momentum $P$ is\be
E(\delta k)=\epsilon_{1}(\delta k)+\epsilon_{2}(-\delta k).
\ee Imposing that $E(\delta k)$ is maximum at $\delta k=0$ implies $u_1=u_2$. That this is a maximum follows from the condition $m_1=m_2<0$.   This means that, for any dispersion relation with a single inflection point above the Fermi level, the upper threshold of the two-particle continuum for $k_F<P<\pi$ has the two particles  in the same subband with negative effective mass. The wave vectors of the particles at the upper threshold are $k_1=k_2=P/2-\pi$ for $k_F<P<\pi/2$, or $k_1=k_2=P/2$ for $\pi/2<P<\pi$ (see Fig. \ref{fig:boundstate}b).

Some intuition about the formation of antibound states can be gained by considering the case $\mathcal{Q}\ll1$, so that the bound state wave function decays slowly and a continuum limit is possible. The Fourier transform of the bound state wave function goes as
\be f(q)\propto {\mathcal{Q}\over q^2+\mathcal{Q}^2}.\ee
The bound state wave function involves relative 
momenta of order $\mathcal{Q}$ or smaller.  Thus it is 
valid to expand the dispersion relation near wave vector $k$ 
when $\mathcal{Q}$ is small. We assume we are in this regime and discuss as example the case $k_1\approx k_2\approx P/2>\pi/2$. We introduce a high energy field $d(x)$  \be
\psi(x)\sim e^{iPx/2}d(x),
\ee 
and expand the dispersion in the form $\epsilon(P/2+p)\approx \varepsilon+up+p^2/2m$ for $p\ll |P|/2$. 
In contrast with the single-particle threshold, we now restrict to the subspace of double occupancy of the $d$ subband\be 
N_d=\int dx\,\bra d^\dagger(x) d(x)\ket=2.\ee It is then important to include in our model the self-interaction of the $d$ field. Neglecting for the moment the interaction with the Fermi surface modes in the case of finite  density, we can write down the effective Hamiltonian density
\be
\mathcal{H}_{pair}=d^\dagger \left(\varepsilon-iu\partial_x-\frac{\partial_x^2}{2m}\right)d^{\phantom\dagger}+\mathcal{V}\,d^\dagger\partial_x d^{\phantom\dagger}\partial_xd^\dagger d^{\phantom\dagger}.\label{Hdtwo}
\ee
In the weak coupling limit, we have $\mathcal{V}\approx V$ and \be\frac{1}{m}\approx\left(1+\frac{V}{\pi}\sin k_F\right) \cos\frac{P}{2}<0.\label{dmass}\ee 
Notice that the interaction between the $d$ particles cannot be described by a delta function potential. This is because for spinless fermions  $s$-wave scattering vanishes identically and the leading process is $p$-wave scattering. In order to obtain a two-particle bound state, we must treat the irrelevant interaction term.

Consider the behavior of the pair spectral function\be
\Pi_{pair}(P,\omega)=\int dx\, e^{-iPx}\int_{-\infty}^{+\infty} dt\, e^{i\omega t}\,\Pi_{pair}(x,t),\ee
\be
\Pi_{pair}(x,t)=\bra d\,\partial_xd(x,t)\partial_xd^\dagger d^\dagger(0,0)\ket,\label{pipair}
\ee
near the upper threshold of the two-particle continuum.
For $\mathcal{V}=0$, the pair spectral function vanishes at the upper threshold of the two-particle continuum as $\Pi_{pair}(P,\omega)\sim\theta(2\varepsilon-\omega)\sqrt{2\varepsilon-\omega}$. For $\mathcal{V}>0$, low-energy scattering between two particles with negative mass can give rise to a bound state above this upper threshold. The wave function for the two  fermionic particles  can be factorized as $\Psi(x_1,x_2)\sim e^{iPX_{cm}}f(r)$, where $X_{cm}=(x_1+x_2)/2$ is the coordinate of the center of mass and  $f(r)$ is the antisymmetric wave function for the relative coordinate $r=x_1-x_2$. This wave function is found by solving the Schr\"odinger equation\be
\left(E-\frac{p^2}{m}\right)f(p)=\int \frac{dq}{2\pi}\,V(q)f(p-q),\label{schrod}
\ee
where $V(q)$ is the Fourier transform of the nearest neighbor interaction potential (in some approximation with a momentum cutoff). 

In one dimension, Eq. (\ref{schrod}) with $m<0$ always gives a normalizable solution with $E>0$ for arbitrarily weak repulsive interaction. However, the wave function of the lowest bound state is an \emph{even} function of the relative coordinate. A second bound state with an antisymmetric wave function can exist for a potential well with finite width and depth if either the interaction is sufficiently strong or the particles are sufficiently heavy. According to Eq. (\ref{dmass}), the effective mass of the $d$ subband is a function of the center-of-mass momentum $P$ and diverges as $P\to\pi$. This corresponds to the momenta of the two particles approaching the inflection point of the dispersion (Fig. \ref{fig:boundstate}b). Therefore, for small $\mathcal{V}>0$, a two-particle bound state with a properly antisymmetric wave function will form when the two particles have total center-of-mass momentum  $P\approx \pi$. Furthermore, the binding energy is maximum at $P=\pi$.  A bound state dispersion $E_{bs}(P)$ can be defined for \be|P-\pi|<\Lambda_{bs},\ee 
where $\Lambda_{bs}\sim O(\mathcal{V})$ is the half-bandwidth of the bound state band. For the integrable model, the exact value of $\Lambda_{bs}$  can be obtained from the BA equations (see Sec. \ref{sec:BA} B).  

The contribution of a bound state to the pair correlation function can be written as \be
\Pi_{pair}(x,t)\sim\left|\frac{df}{dr}(0)\right|^2\sum_Pe^{-iPx-iE_{bs}(P)t}.
\ee
In analogy with the definition of $d$ subbands for the single-particle dispersion, we take the continuum limit in the bound state dispersion and introduce a subband with momentum $P\approx P_0$, with $|P_0-\pi|<\Lambda_{bs}$. We expand\be
P\approx P_0+p\quad,\quad \mathcal{E}(P)\approx \varepsilon_{bs}+u_{bs}p+p^2/2m_{bs},
\ee
where $\varepsilon_{bs}= \mathcal{E}(P_0)$ and $u_{bs}$ and $m_{bs}$ are the effective bound state velocity and mass around momentum $P_0$. Since the bound state dispersion has a minimum at $P=\pi$, we have $m_{bs}>0$. The propagator for a high energy bound state  is \be
\Pi_{pair}(x,t)\approx \left|\frac{df}{dr}(0)\right|^2\,e^{-i\varepsilon_{bs}t}\delta(x-u_{bs}t).\label{proppair}
\ee

We now turn to the calculation of the contribution of the two-bound-particle one-hole state to the single-particle Green's function for $k_F<k<\pi$. As explained in Ref. \onlinecite{khodas}, the Schrieffer-Wolff transformation that eliminates from the Hamiltonian off-diagonal scattering processes (those which take particles out of their subbands) generates higher-order contributions to the fermionic field. For instance, the leading contribution to $G_p(k,t)$ due to a state with  two particles in the same high-energy subband and  one hole at the right Fermi point is given by\be
G_p(k,t)=\int dx\,e^{-ikx}\bra\psi(x,t)\psi^\dagger(0,0) \ket,
\ee 
with\bea
\psi^\dagger(x)&\sim& e^{-ikx}d^\dagger (x)\partial_xd^\dagger(x)\psi_R(x)\label{projectB}.
\eea
The amplitude on the rhs of  Eq. (\ref{projectB}) is $O(V)$. The idea now is that, if the total wave vector $k$ is in the range where the two high-energy particles can bind,  the two $d^\dagger$ operators in Eq.  (\ref{projectB}) can be replaced  by a single operator corresponding to the creation of  a bound state. We assume this to be valid both when $\mathcal{Q}$ is small and when $\mathcal{Q}$ is $O(1)$ (tightly bound pair). 

The minimum momentum $k_b$ for which a bound state contribution to $G_p(k,t)$ exists  is obtained by having the bound state with the lowest allowed momentum $P_0=\pi-\Lambda_{bs}$ and the hole with the maximum allowed momentum $k_F$:\be
k_b=\pi-\Lambda_{bs}-k_F.
\ee
For $k>k_b$, there is a continuum of states defined by a single pair and a single hole. For the integrable model, the energy of an excited state with total momentum $k$ can be written as\be
E=\mathcal{E}(P_0)-\varepsilon(P_0-k),\label{energypairhole}
\ee
where $\mathcal{E}(P_0)$ and $\varepsilon(P_0-k)$ are the renormalized energies of the bound state and hole, respectively. 

The lower threshold of the one-bound-state one-hole continuum is determined by minimizing  Eq. (\ref{energypairhole}) with respect to $P_0$ for $|P_0-\pi|\leq\Lambda_{bs}$ and $|P_0-k|\leq k_F$. There are three possibilities. 

a) First, for $u_{bs}(k+k_F)<v$, the lower threshold is given by a hole at $k_F$ and a bound state with momentum $P_0=k+k_F$. In this case, we consider the correlation function for the operator in Eq. (\ref{projectB}). We define the pair creation operator\be
B^\dagger(x)=\frac{1}{\sqrt{2}}\int dr\, f(r)\,d^\dagger\left(x+\frac{r}{2}\right)d^\dagger\left(x-\frac{r}{2}\right).
\ee
The $B^\dagger$ field is bosonic (it creates a $p$-wave ``molecule"). We  project the fermionic field into the subspace of a single bound state  \be
\psi^\dagger(x)\sim e^{-ikx}B^\dagger(x)\psi_R(x).
\ee
We can now  write down an effective Hamiltonian which phenomenologically describes the coupling of the bound state to the low-energy modes  \bea
\mathcal{H}_2&=&\nonumber B^\dagger(\varepsilon_{bs}-iu_{bs}\partial_x)B+\frac{v}{2}\left[\left(\partial_x\varphi_R\right)^2+\left(\partial_x\varphi_L\right)^2\right]\nonumber\\
&&+\frac{1}{\sqrt{2\pi K}}\left(\kappa^{bs}_{L}\partial_x\varphi_L-\kappa^{bs}_{R}\partial_x\varphi_R\right)B^\dagger B^{\phantom\dagger}.\label{couplebsboson}\eea
For the integrable model, all the coupling constants can be extracted from the BA as discussed in Sec. \ref{sec:BA} B.
The interaction in Eq. (\ref{couplebsboson}) can be diagonalized by the unitary transformation of Eq. (\ref{unitary}) with the appropriate parameters \be
\gamma_{R,L}^{bs}=\frac{\kappa^{bs}_{R,L}}{v\mp u_{bs}}.\ee
The contribution to the particle Green's function becomes\be
G_p(k,t)\sim \int dx\, \frac{\Pi_{pair}(x,t)}{(x-vt+i\eta)^{\nu_R^\prime}(x+vt-i\eta)^{\nu_L^\prime}},\label{Gboundu<v}
\ee
where \be
\nu^\prime_{R,L}(k)=\frac{1}{4K}\left(1-\frac{\gamma^{bs}_{R,L}}{\pi}\pm K\right)^2,\label{nusbound}
\ee 
and $\Pi_{pair}(x,t)$ is the bound state part of the pair correlation function given in Eq. (\ref{proppair}). As a result, in the case $u_{bs}(P_0=k+k_F)<v$, the contribution to $G_p(k,t)$ is given by\be
G_p(k,t)\sim e^{-i\varepsilon_{bs}t}(t-i\eta)^{-\nu_R^\prime-\nu_L^\prime}.
\ee
Since $\gamma_{R}^{bs}>0$ for $V>0$, one can verify in the weak coupling limit that $\nu_R^\prime+\nu_L^\prime<1$ and $1-\nu_R^\prime-\nu_L^\prime\sim O(V)$. As $G_p(k,t)$ decays with an exponent smaller than 1, the spectral function acquires a divergent edge singularity \be
A_p(k,t)\sim\theta(\omega-\varepsilon_{bs})(\omega-\varepsilon_{bs})^{\nu_R^\prime+\nu_L^\prime-1}.\label{Apboundweak}
\ee
This divergence in the spectral function can be interpreted as an x-ray edge singularity due to the repulsive interaction between the bound state and the particles at the Fermi surface.

If $V$ is small, the bound state band is narrow and we expect   $u_{bs}(P_0)<v$ for all $\pi-\Lambda_{bs}<P_0<\pi+\Lambda_{bs}$. In this case, the singularity in Eq. (\ref{Apboundweak}) is present for all  $\pi-\Lambda_{bs}-k_F<k<\pi+\Lambda_{bs}-k_F$.

b) The second possibility when minimizing Eq. (\ref{energypairhole})  is that, for $u_{bs}\left(\pi+\Lambda_{bs}\right)<v$ and $k>\pi+\Lambda_{bs}-k_F$, it is still possible to create one-bound-state one-hole excitations by creating the hole at a state with momentum $k_h=P_0-k<k_F$. In this case, the lower threshold of the continuum  has three high-energy particles.  For small values of $k_F-k_h$, the lower threshold is defined by a bound state with the maximum momentum $P_0=\pi+\Lambda_{bs}$ and a deep hole with momentum $k_h=\pi+\Lambda_{bs}-k$, such that the velocity of the deep hole is $u_h(k_h)>u_{bs}(\pi+\Lambda_{bs})$. 

As we keep increasing $k$ and decreasing $k_h$, there will be a value of hole momentum $k_h^*$ at which the hole velocity $u_h(k_h^*)$ equals the maximum bound state velocity $u_{bs}(P_0=\pi+\Lambda_{bs})$. For $k>\pi+\Lambda_{bs}-k_h^*$, minimizing Eq. (\ref{energypairhole}) gives a lower threshold  defined by a bound state and a deep hole with equal velocities: $u_h(P_0-k)=u_{bs}(P_0)<v$. 

The behavior of the particle Green's function near the lower threshold with a bound state and a deep hole can be calculated using\be
\psi^\dagger(x)\sim e^{-ikx}B^\dagger(x)d_h^\dagger(x),\label{B+hole}
\ee
where $d_h^\dagger(x)$ creates a hole. The operators in Eq. (\ref{B+hole}) must be defined after solving the three-body problem in which the two particles effectively attract each other (due to the negative mass) and repel the extra hole. We assume that this has been done and the resulting spectrum of the one-bound-state one-free-particle type excitations has been parametrized as in Eq. (\ref{energypairhole}), with no residual interactions between the high-energy modes. The coupling of these high energy particles to the low-energy modes can be described by the effective Hamiltonian density\bea
\mathcal{H}_3&=&B^\dagger \left(\varepsilon_{bs}-iu_{bs}\partial_x-\frac{\partial_x^2}{2m_{bs}}\right)B\nonumber\\
&&+d_h^\dagger \left(\varepsilon_h-iu_h\partial_x-\frac{\partial_x^2}{2m_h}\right)d_h^{\phantom\dagger}\nonumber\\
&&+\frac{1}{\sqrt{2\pi K}}\left(\kappa^{bs}_{L}\partial_x\varphi_L-\kappa^{bs}_{R}\partial_x\varphi_R\right)B^\dagger B^{\phantom\dagger}\nonumber\\
&&+\frac{1}{\sqrt{2\pi K}}\left(\kappa^{h}_{L}\partial_x\varphi_L-\kappa^{h}_{R}\partial_x\varphi_R\right)d_h^\dagger d_h^{\phantom\dagger}.\label{H3part}\eea
The particle Green's function derived from Eqs. (\ref{B+hole}) and (\ref{H3part}) is\be
G_p(k,t)\sim \int dx \frac{\Pi_{pair}(x,t)G^{(0)}_h(x,t)}{(x-vt+i\eta)^{\nu_R^{\prime\prime}}(x+vt-i\eta)^{\nu_L^{\prime\prime}}},\label{Gp3part}
\ee
where $G^{(0)}_h(x,t)$ is the propagator of the free deep hole and $\nu_{R,L}^{\prime\prime}\sim O(V^2)$ are given by Eq. (\ref{nus}) with the total phase shifts $\gamma_{R,L}=\gamma^{bs}_{R,L}+\gamma^h_{R,L}$ due to the bound state as well as the deep hole. 

One can verify that, as long as the velocity of the deep hole is larger than the velocity  of the bound state, Eq. (\ref{Gp3part}) does not lead to a divergence at the lower threshold of the one-bound-state one-deep-hole continuum. Therefore, for $u_{bs}(\pi+\Lambda_{bs})<v$  there is no divergent singularity in the wave vector range $\pi+\Lambda_{bs}-k_F<k<\pi+\Lambda_{bs}-k_h^*$. 

For $k>\pi+\Lambda_{bs}-k_h^*$, we must set $u_h=u_{bs}$ in Eq. (\ref{H3part}). In this case, replacing both the hole and bound state propagators by delta functions with the same argument as in Eqs. (\ref{propfreepart}) and (\ref{proppair}) would lead to a divergence. We must instead calculate these propagators by keeping the quadratic term in each dispersion relation. This gives\bea
\Pi_{pair}(x,t)&\sim&e^{-i\varepsilon_{bs}t}\sqrt{\frac{m_{bs}}{2\pi it}}e^{im_{bs}(x-u_{bs}t)^2/2t},\\
G_h(x,t)&\sim&e^{-i\varepsilon_{h}t}\sqrt{\frac{m_{h}}{2\pi it}}e^{im_{h}(x-u_{bs}t)^2/2t}.
\eea
The integral in Eq. (\ref{Gp3part}) is then dominated by the region $x\approx u_{bs}t$. In the limit of large $t$, we obtain\be
G_p(k,t)\sim \sqrt{\frac{m_{bs}m_h}{2\pi it(m_{bs}+m_h)}}\,\frac{e^{-i(\varepsilon_{bs}+\varepsilon_h)t}}{(t-i\eta)^{\nu_R^{\prime\prime}+\nu_L^{\prime\prime}}}.
\ee
This leads to a divergence in the spectral function\be
A(k,\omega)\sim\theta(\omega-\varepsilon_{bs}-\varepsilon_h)(\omega-\varepsilon_{bs}-\varepsilon_h)^{-\frac{1}{2}+\nu_R^{\prime\prime}+\nu_L^{\prime\prime}}. \label{Apboundstrong}
\ee
In the weak coupling limit, this is essentially the divergence of the joint density of states of the two-particle bound state and hole with the same velocity. 

In summary, in the case $u_{bs}(\pi+\Lambda_{bs})<v$, the divergent edge singularity due to  a two-particle bound state is absent for $\pi+\Lambda_{bs}-k_F<k<\pi-\Lambda_{bs}-k_h^*$ but resurges for $\pi-\Lambda_{bs}-k_h^*<k<\pi$ as a stronger singularity, with exponent close to $-1/2$ for  $V\ll1$.

c) Finally, the third possibility, when minimizing Eq. (\ref{energypairhole}), is that $u_{bs}(P_0)>v$ for allowed values of $P_0<\pi+\Lambda_{bs}$. This condition is more likely to be observed at  lower densities (smaller $v$) and stronger interactions (larger $\Lambda_{bs}$). In this case, a divergent edge singularity exists for all $k>k_b$. The nature of the lower threshold changes at $k=P_0^*-k_F$, where $P_0^*$ is such that $u_{bs}(P_0^*)=v$. For $\pi-\Lambda_{bs}-k_F<k<P_0^*-k_F$, the lower threshold is defined by a hole at $k_F$ and a  bound state with momentum $k+k_F$. The corresponding  singularity is the weaker one given by Eq. (\ref{Apboundweak}). For $P_0^*-k_F<k<\pi$, the lower threshold is defined by a bound state and a deep hole with the same velocity. The edge singularity in this case is described by Eq. (\ref{Apboundstrong}).

\section{Integrability and power-law singularities\label{sec:integrable}}
\subsection{Broadening of the  single-particle peak for $u>v$\label{sec:decaysmallk}}
The result in Sec. \ref{sec:ham} predicts that the single-particle spectral function exhibits  a divergent power-law singularity at the energy of the single-particle excitation. The reason for the exact power-law decay of the particle Green's function can be traced back to the ballistic propagation of the high-energy particle, as expressed by the free propagator in Eq. (\ref{propfreepart}). If the propagation of the $d$ particle is made diffusive, in the sense of a nonzero decay rate $1/\tau$ in the particle Green's function, the divergent singularities in the spectral functions are replaced by rounded, Lorentzian-like peaks. In real time, the Green's function decays exponentially as $G_p(k,t)\sim e^{-t/\tau}$ at large $t$.\cite{khodas}

The purpose of this section is to argue that a nonzero decay rate $1/\tau$ in the regime $u>v$  (``fast particle") can be recovered within the effective  field theory approach by adding to Eq. (\ref{Hdecoupled}) the following irrelevant interaction term which couples the $\bar d$ particle to the bosonic fields \be
\delta\mathcal{H}=g\,\bar{d}^{\dagger}\bar{d}^{\phantom\dagger}\partial_x\bar\varphi_R\partial_x\bar\varphi_L,\label{deltaHdecay}
\ee
where $g$ is the coupling constant. In the original fermion
representation, this corresponds to a six-fermion interaction term: $\bar
d^\dagger \bar d\psi_R^\dagger \psi^{\phantom\dagger}_R\psi_L^\dagger \psi^{\phantom\dagger}_L$. It leads to
a three-body scattering process where the high energy $d$ particle scatters
simultaneously off two particles slightly below the Fermi energy (one near
the left branch and one near the right branch), producing a final state
with the momentum of the $d$ particle slightly shifted and two particles
produced slightly above the Fermi energy  (one near the left branch and one
near the right branch). The same interaction was considered in Ref. \onlinecite{castroneto}, where it was shown to give rise to the finite scattering rate of a mobile impurity coupled to a Luttinger liquid.  This type of interaction is allowed by symmetry and is generated by bosonization of the interaction term in Eq. (\ref{model}) to second order in $V$. It is also generated by the unitary transformation that eliminates the marginal coupling between $d$ and the bosonic fields. However, since $g$ is $O(V^2)$, the correct $g$ cannot be fixed in the weak coupling limit by simple bosonization. 

Let us assume that $g$ is present in the effective Hamiltonian and look at the consequences.
Consider the effective Hamiltonian for a single high-energy particle with momentum near $k$. From Eq. (\ref{Hdecoupled}), the time-ordered free propagator for the $d$ particle is\be
G^{(0)}(x,t)=\bra T \bar{d}(x,t)\bar{d}^\dagger(0,0)\ket=\theta(t)e^{-i\varepsilon t}\delta(x-ut).
\ee
The Fourier transform is\be
G^{(0)}(p,\omega)=(\omega-\varepsilon-up+i\eta)^{-1},
\ee
where $|p|<\eta^{-1}\ll k-k_F$ is the momentum within the subband. We apply perturbation theory in the $g$ interaction. The Dyson equation for the $d$ propagator gives the dressed propagator \be
G(p,\omega)=[\omega-\varepsilon-up-\Sigma(p,\omega)]^{-1},
\ee
where $\Sigma(p,\omega)$ is the self-energy. To second order in the coupling constant, we have\bea
\Sigma(p,\omega)&=&-ig^2\int_{-\infty}^{\infty} dx\,e^{-ipx}\int_{-\infty}^{\infty} dt \,e^{i\omega t}D^{(0)}_R(x,t)\nonumber\\&&\times D^{(0)}_L(x,t)G^{(0)}(x,t),
\eea
where\bea
D^{(0)}_{R,L}(x,t)&\equiv&\bra\partial_x\bar\varphi_{R,L}(x,t)\partial_x\bar\varphi_{R,L}(0,0)\ket_0\nonumber\\
&=&-\frac{1}{2\pi(vt-i\eta \mp x)^2}.
\eea
The imaginary part of the self-energy reads\bea
\textrm{Im }\Sigma(p,\omega)&=&-\left(\frac{g}{2\pi}\right)^2 \int_{-\infty}^{\infty}dt\,\nonumber\\
&&\times\frac{e^{i(\omega-\varepsilon-up)t}}{[(u-v)t+i\eta]^2[(u+v)t-i\eta]^2}.\label{imagsigma}\eea
We can see that,  if the velocity of the high-energy particle is larger than the Fermi velocity, the integrand in Eq. (\ref{imagsigma}) has poles both below and above the real axis and Im $\Sigma$ is nonzero. This is precisely the condition for the existence of the two-sided singularity at the single-particle energy. In contrast, if $d$ is a deep hole or a high energy particle with $u<v$, Im $\Sigma$ vanishes identically. Note also that interaction  processes such as $\bar{d}^{\dagger}\bar{d}^{\phantom\dagger}(\partial_x\bar\varphi_R)^2$ or $\bar{d}^{\dagger}\bar{d}^{\phantom\dagger}(\partial_x\bar\varphi_L)^2$, in which the $\bar{d}$ particle scatters only right movers or only left movers, do not contribute to the imaginary part of the self-energy.

Assuming $u>v$, the decay rate for the high-energy particle with momentum near $k$ is\be
\frac{1}{\tau}=-\textrm{Im }\Sigma(p,\omega=\varepsilon+up)=\frac{g^2(u^2-v^2)}{\pi(2u\eta)^3},\label{decayrate}
\ee
where $\eta$ is the short distance cutoff of the bosonic fields. Therefore, this approach yields a cutoff-dependent yet nonzero decay rate. It is instructive to consider the scaling of the decay rate in the low-energy limit $k\to k_F$. In this limit, we have\be
u-v\approx (k-k_F)/m.
\ee
Moreover, the cutoff must scale with the separation between the low and high-energy subbands, therefore\be
\eta^{-1}\sim C|k-k_F|,
\ee
where $C\ll1$ is a constant. Eq. (\ref{decayrate}) then simplifies to\be
\frac{1}{\tau}\sim\frac{C g^2(k-k_F)^4}{\pi m(2v)^2}.\label{tausimpler}
\ee
Finally, in the weak coupling limit $g$ must be regarded as the amplitude for a decay process in which the high energy particle scatters one right mover and one left mover. In terms of the original fermions, the operator in Eq. (\ref{deltaHdecay}) couples a single-particle state to a state with three particles and two holes\be
|\alpha\ket=\psi^{\dagger}_{k_1}\psi^{\dagger}_{k_2}\psi^{\dagger}_{k_3}\psi^{\phantom\dagger}_{k_4}\psi^{\phantom\dagger}_{k_5}|0\ket.\label{alpha3}
\ee
For a generic two-body interaction potential, $g\sim\bra\alpha|\psi^\dagger_k|0\ket$ must be $O(V^2)$ and must vanish in the limit $k\to k_F$ (no $s$-wave scattering for spinless fermions). Furthermore, the coupling constant $g$ has dimensions of $1/m$. The combination of parameters that satisfies these conditions is the one derived in Ref. \onlinecite{khodas} by perturbation theory\be
g\propto \frac{V_0(V_0-V_{k-k_F})}{mv^2}.
\ee
Expanding $V_q\approx V_0+V_0^{\prime\prime}q^2/2$ for a short-range interaction and substituing this form into Eq. (\ref{tausimpler}), we find\be
\frac{1}{\tau}\propto\frac{(V_0V^{\prime\prime}_0)^2(k-k_F)^8}{\pi m^3v^6}.\label{decayratelowen}
\ee

Eq. (\ref{decayratelowen}) recovers the interaction and momentum dependence of the decay rate found in Ref. \onlinecite{khodas} for small $V$ and $k\approx k_F$. In this limit, the decay rate is extremely small. However,  based on the result of Eq. (\ref{decayrate}), we expect that the decay rate for  $V$ and $k-k_F$ of order 1 can be fairly large in a generic model. This would imply that, at high energies (but still in the regime $u>v$) and for strong interaction, the particle spectral function for a generic (nonintegrable) 1D model exhibits a broad peak near the single-particle energy, not much different from the higher-dimensional counterpart. 

However, for an integrable model, it is possible (and perhaps probable) that $g$ vanishes identically, since it is  related to the amplitude of a three-body collision, in the sense discussed below Eq. (\ref{deltaHdecay}). Thus we conjecture that the irrelevant operator in Eq. (\ref{deltaHdecay}) is absent in the effective Hamiltonian for an integrable model. In fact, the exact vanishing of the decay rate to all orders in $V$ and $k-k_F$ requires the absence of infinitely many more irrelevant interaction terms which couple the $\bar{d}$ particle to the bosonic fields and involve higher powers of $\partial_x\bar\varphi_{R,L}$. This is only possible by fine tuning of the coupling constants. That integrability poses nontrivial constraints on the coupling constants of some irrelevant operators has been demonstrated rigorously in Ref. \onlinecite{pereiraJSTAT}. Here we shall justify our conjecture in Sec. \ref{sec:DMRG} by examining the numerical tDMRG results for the particle Green's function of the integrable model (\ref{model}).

\subsection{Broadening of the single-particle peak for $u<v$\label{sec:decaylargek}}

In the ``slow particle" regime, where the velocity of the high-energy particle is smaller than the Fermi velocity, the three-body collision type interaction discussed in the previous section yields zero decay rate. This regime does not occur in the continuum model.\cite{khodas} The motivation to consider three-body collision type processes for $u>v$ was that in the continuum model the decay rate vanishes at second order in the interaction, making it necessary to go to $O(V^4)$. In this section we point out that in the lattice model the decay rate for a single-particle excitation with  $u<v$ is nonzero at $O(V^2)$. This is due a two-body collision process in which the slow particle scatters off one particle deep below the Fermi surface and both final particles are, in general, at high energy states above the Fermi surface.  Therefore, the broadening of the single-particle peak for $u<v$ does not require three-body collisions and is  present even in the integrable model.

\begin{figure}
\includegraphics*[width=0.5\hsize,scale=1.0]{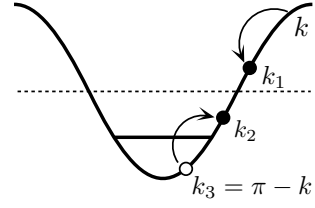}
\caption{Finite-momentum decay process which gives a finite  decay rate for the high-energy particle with $u<v$. \label{fig:decayulessv}}
\end{figure}

We calculate the decay rate in the limit $V\ll1$ by straightforward perturbation theory in the interaction. We write the Hamiltonian in the form $H=H_0+H_{int}$, where $H_0$ is given by Eq. (\ref{Hnonint}) and  \be
H_{int}=\frac{1}{2L}\sum_{k_1,k_2,q}V(q)\psi^\dagger_{k_1} \psi^{\phantom\dagger}_{k_1+q} \psi^\dagger_{k_2}\psi^{\phantom\dagger}_{k_2-q}.
\ee
It follows from Fermi's golden rule that the imaginary part of the self-energy to $O(V^2)$ is\bea
\textrm{Im }\Sigma^{(2)}(k,\omega)&=&-\frac{\pi}{2}\sum_{p,q\neq0}[V(q)-V(k-p-q)]^2\nonumber\\
&&\times\theta(\epsilon^{(0)}_{k-q})\theta(-\epsilon^{(0)}_{p})\theta(\epsilon^{(0)}_{p+q})\nonumber\\
&&\times\delta(\omega-\epsilon^{(0)}_{k-q}-\epsilon^{(0)}_{p+q}+\epsilon^{(0)}_{p}).\label{selfenergy(2)}
\eea
In contrast with Eq. (\ref{alpha3}), the final state $|\alpha\ket$ considered in Eq. (\ref{selfenergy(2)}) has only two particles above the Fermi level (in general both of them at high energies) and the amplitude $\bra\alpha|\psi_k^\dagger|0\ket$ is $O(V)$.  If the dispersion is parabolic, Im $\Sigma^{(2)}$ vanishes at the single-particle energy due to phase space constraints. However, if the dispersion has an inflection point, it is possible to satisfy both momentum and energy conservation such that the decay rate is nonzero. In order to calculate $1/\tau$ to $O(V^2)$, we can use the noninteracting dispersion $\epsilon^{(0)}_k=-2\cos k-\mu$. Thus the ``on-shell" condition \be
\epsilon^{(0)}_k+\epsilon^{(0)}_p-\epsilon^{(0)}_{k-q}-\epsilon^{(0)}_{p+q}=0\label{onshell}
\ee
is satisfied for\be
 p=\pi-k\quad,\quad k+k_F-\pi<q<k-k_F.
 \ee
This decay process is only kinematically allowed for $k>\pi-k_F$, the momentum range where $u>v$ in the weak coupling limit.   It involves  a finite-momentum scattering process in which the high-energy particle decays as a hole is created in the state with momentum $\pi-k$ and another high-energy particle is created above the Fermi surface (see Fig. \ref{fig:decayulessv}). There is a continuum of two-particle one-hole excitations, corresponding to different choices of $q$, which are degenerate with the single-particle excitation. Importantly, the allowed range of $q$ shrinks to zero in the limit $k_F\to\pi/2$ and the decay rate  vanishes at half-filling. 

\begin{figure}
\includegraphics*[width=0.9\hsize,scale=1.0]{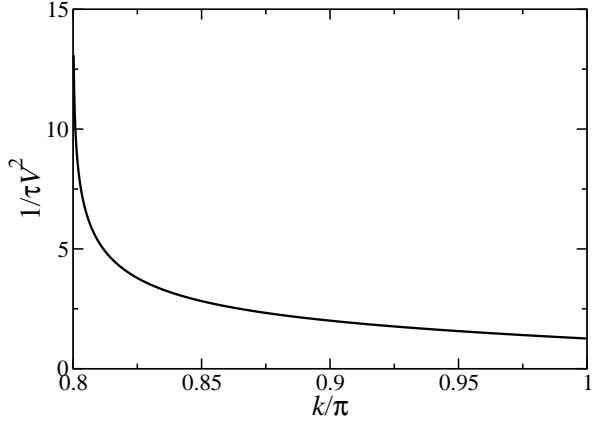}
\caption{Second-order result for the decay rate of the high-energy particle in the wave vector range  $k>\pi-k_F$, for $k_F=\pi/5$. \label{fig:decayrate}}
\end{figure}

Substituting $V(q)=2V\cos q$ for the nearest neighbor interaction potential  in Eq. (\ref{selfenergy(2)}), we find the decay rate\be
\frac{1}{\tau}=2V^2\cos^2 k\int_{k_F}^{\pi-k_F}dk_1\,\frac{\cos^2 k_1}{\sin k_1-\sin k}.
\ee
The explicit result is cumbersome, but the behavior of $1/\tau$ as a function of $k$ is illustrated in Fig. \ref{fig:decayrate} for $k_F=\pi/5$. There is a logarithmic divergence at $k\to\pi-k_F$. This is an artifact of stopping at second order in perturbation theory. The decay process for $k\to\pi-k_F$ involves the creation of a hole at the Fermi point with momentum $k_F$. We expect that at higher orders  in $V$ the density of states at the Fermi surface should acquire a power-law suppression due to Luttinger liquid physics. As a result,  $1/\tau$ should vanish as $k$ approaches the lower threshold.  A similar discontinuity in the decay rate which is removed by treating interactions in the final state  has been discussed in the context of magnons in the Haldane chain in a magnetic field.\cite{essler}

The kinematic boundary for this decay process can be generalized for finite interaction strength $V$. Assuming that the renormalized dispersion has a single inflection point above $k_F$, the minimum value of $k$ for which $1/\tau>0$ is obtained from the condition that Eq. (\ref{onshell})  be satisfied for the hole at the Fermi point with momentum $k_F$\be
\epsilon_k-\epsilon_{k-q}-\epsilon_{k_F+q}=0\label{onshell2}.
\ee 
Imposing that Eq. (\ref{onshell2}) is satisfied for $q\to0$ leads to the condition $u=v$. We conclude that the decay rate is nonzero and the spectral function has a broadened single-particle peak for momentum $k$ in the regime $u<v$. The precise momentum and interaction dependence of the decay rate in this regime is an open question.

In real time, the corresponding contribution to the particle Green's function  decays exponentially at large $t$:\cite{khodas}\be
G_p(k,t)\sim \frac{\textrm{exp}(-i\varepsilon t-i\pi\nu/2-t/\tau)}{t^{\nu}}.\label{Gexpdecay}
\ee

\section{Exact energies and exponents from the Bethe ansatz\label{sec:BA}}
In order to fix the parameters of the effective model in Eq. (\ref{Hafterscaling}), we need a BA calculation of a physical quantity which depends on the coupling constants $\kappa_{R,L}$ or, equivalently, $\gamma_{R,L}$. First, let us show that the field theory approach predicts that the finite size spectrum for low-energy excitations in the presence of a high-energy particle depends on these coupling constants. Following Ref. \onlinecite{eggert}, we can show that the finite size spectrum for the Luttinger model (without the high-energy particle) with periodic boundary conditions is \be
\Delta E=\frac{2\pi v}{L}\left(\frac{\Delta N^2}{4K}+KD^2+n_++n_-\right),\label{spectrum}
\ee
where $\Delta N$ is integer, $D$ is integer (half-integer) for $N$ even (odd) and $n_\pm$ are integers. $\Delta N$ measures the number of low energy charge excitations.  We have\bea
\Delta N&=&\int_{0}^L dx\, \bra\psi^\dagger_R\psi^{\phantom\dagger}_R+\psi^\dagger_L\psi^{\phantom\dagger}_L\ket\nonumber\\
&=&\sqrt{\frac{K}{2\pi}}\int_0^Ldx\,\bra\partial_x\varphi_L-\partial_x\varphi_R\ket.
\eea
Changing the total number of particles changes the boundary conditions of the bosonic field in a finite size system.\cite{eggert} Likewise, $D$ is the number of low-energy current excitations, in which particles are transferred between the two Fermi points. Hence\bea
D&=&\frac{1}{2}\int_0^L dx\, \bra\psi^\dagger_R\psi^{\phantom\dagger}_R-\psi^\dagger_L\psi^{\phantom\dagger}_L\ket\nonumber\\
&=&-\frac{1}{\sqrt{8\pi K}}\int_0^Ldx\,\bra\partial_x\varphi_L+\partial_x\varphi_R\ket.
\eea
In the presence of a single $d$ particle ($\bra d^\dagger d\ket=1/L$), the unitary transformation of Eqs. (\ref{unitary}) and (\ref{phishift}) takes \bea
\Delta N&\to&\Delta N-\frac{\gamma_R+\gamma_L}{2\pi},\\
D&\to&D-\frac{\gamma_R-\gamma_L}{4\pi K}.
\eea
The spectrum of the low-energy particle-hole excitations about the Fermi points ($n_\pm$ terms in Eq. (\ref{spectrum})) is not modified by the creation of the high-energy particle. We conclude that the finite size spectrum in the presence of the $d$ particle (including the term of $O(1)$ from Eq. (\ref{Hdecoupled})) must assume the form\bea
\Delta E&=&\varepsilon+\frac{2\pi v}{L}\left[\frac{1}{4K}\left(\Delta N-\frac{\gamma_R+\gamma_L}{2\pi}\right)^2\right.\nonumber\\&&\left.+K\left(D-\frac{\gamma_R-\gamma_L}{4\pi K}\right)^2+n_++n_-\right].\label{spectrumshift}
\eea
This is the spectrum of a shifted $c=1$ conformal field theory.\cite{kawakami} The parameters $\gamma_{R,L}$, which determine the exponents in Eq. (\ref{nus}), can be interpreted as renormalized phase shifts at the Fermi points due to the creation of the $d$ particle.

We now proceed to calculate the finite size spectrum in the BA, following Woynarovich.\cite{woynarovich} We should first point out that there is no correspondence between   the fermions of the effective field theory and the quasiparticles of BA eigenstates. However, we can reasonably expect that the power-law singularities of the spectral function, if existing,  will develop at the energy thresholds of the exact spectrum. We can focus on a particular (in practice, the simplest possible) class of eigenstates, compute its finite size spectrum and compare it with the expression in Eq. (\ref{spectrumshift}). This procedure does not require assumptions about the form factors of states that contribute to the spectral function as written in Eqs. (\ref{Apart}) and (\ref{Ahole}). 

\subsection{Single particle excitations}

We start from the Bethe equations for the eigenstates of the Hamiltonian in Eq. (\ref{model})  with $N<L/2$ fermions  \cite{korepin}\begin{equation}
Lp_{0}\left(\lambda_{j}\right)=2\pi I_{j}+\sum_{k=1}^{N}\theta\left(\lambda_{j}-\lambda_{k}\right).\label{BAequation}\end{equation}
The quantum numbers $I_{j}$ are half-integers for $N$ even and integers
for $N$ odd. The functions $p_{0}(\lambda)$ and $\theta(\lambda)$
are, respectively, the quasimomentum and two-particle scattering phase
shift as a function of rapidity $\lambda$ \begin{eqnarray}
p_{0}\left(\lambda\right) & = & i\log\left[\frac{\sinh\left(i\zeta/2+\lambda\right)}{\sinh\left(i\zeta/2-\lambda\right)}\right],\label{p0}\\
\theta\left(\lambda\right) & = & i\log\left[\frac{\sinh\left(i\zeta+\lambda\right)}{\sinh\left(i\zeta-\lambda\right)}\right],\label{theta}\end{eqnarray}
where $\zeta=\arccos(V/2)$. A given eigenstate is characterized by a set  of rapidities $\{\lambda_{j}\}$ and the corresponding energy is
\be
E=\sum_{j=1}^{N}\epsilon_{0}\left(\lambda_{j}\right),\ee
where\be
\epsilon_{0}\left(\lambda\right)=-\frac{2\sin^{2}\zeta}{\cosh(2\lambda)-\cos\zeta}-\mu\label{bareenergy}\ee
is the bare energy of a particle with rapidity $\lambda$.

Consider a state defined by taking $\{ I_{j}\}$ to be the set of all integers (or half-integers) between $I_{+}$ and $I_{-}$, which are defined by\begin{eqnarray}
I_{+} & = & \max\left\{ I_{j}\right\} +\frac{1}{2},\\
I_{-} & = & \min\left\{ I_{j}\right\} -\frac{1}{2}.\end{eqnarray}
This is a state with no low-energy particle-hole pairs ($n_{\pm}=0$). The total number of particles is\begin{equation}
N=I_{N}-I_{1}+1=I_{+}-I_{-}.\label{eq:MIplusIminus}\end{equation}
The current carried by an eigenstate is given by the difference between the number of right and left movers. It can be written as \begin{equation}
2D=I_{1}+I_{N}=I_{+}+I_{-}.\label{eq:DIplusIminus}\end{equation}

The ground state for $N$ particles corresponds to the choice $I_{+}=-I_{-}=N/2$, with all rapidities $\lambda_j$ in Eq. (\ref{BAequation}) real.   To create a single-hole excitation we remove a quantum number $I_h$, $|I_h|<N/2$, from the set of $I_j$'s.  A single particle excitation is created by adding one quantum number $I_p$ with $|I_p|>N/2$. We distinguish between single-particle excitations with real and complex rapidities. It follows from Eq. (\ref{BAequation}) that, for real rapidities $|\lambda_j|<\infty$, the quantum numbers $I_j$ are restricted to the interval $N/2<|I_j|<I_\infty$, where\cite{pereiraJSTAT}\be
I_\infty=\frac{L-N}{2}-\frac{\pi-\zeta}{\pi}\left(\frac{L}{2}-N\right).
\ee 
As we shall see, this implies a maximum momentum for single particle excitations with real rapidities. Note, however, that we also get real values of  $p_0(\lambda)$ and $\theta(\lambda)$ in Eqs. (\ref{p0}) and (\ref{theta}) (therefore a scattering  state of unbound quasiparticles) by taking complex rapidities of the form\be
\lambda_p=\textrm{ Re}(\lambda_p)+\frac{i\pi}{2}.\label{lambdastring}
\ee
This is called a negative-parity one-string.\cite{caux} This type of solution has quantum numbers  in the interval $\tilde{I}_\infty<|I_p|<L/2$, where\be
\tilde{I}_\infty=\frac{N}{2}+\frac{\pi-\zeta}{\pi}\left(\frac{L}{2}-N\right).
\ee  

In the limit of large $L$, Eq. (\ref{BAequation}) becomes an equation for the density of rapidities $\rho(\lambda)$\cite{korepin}
\begin{equation}
\rho\left(\lambda_{j}\right)=\frac{1}{2\pi}\left[p_{0}^{\prime}\left(\lambda_{j}\right)+\frac{1}{L}\sum_{k}K\left(\lambda_{j}-\lambda_{k}\right)+\frac{\Phi_{p,h}\left(\lambda_{j}\right)}{L}\right].\label{eq:BAeqfsfinal}\end{equation}
Here the sum is over all the quantum numbers between $I_{-}$ and $I_{+}$. The kernel $K(\lambda)$ is given by\be
K(\lambda)=-\frac{d\theta(\lambda)}{d\lambda}=\frac{2\sin(2\zeta)}{1-2\cos^2\zeta+\cosh(2\lambda)}.
\ee  
The last term on the rhs of Eq. (\ref{eq:BAeqfsfinal}) is the  scattering phase shift between the particle at $\lambda_j$ and the high-energy particle or hole. In the case of a hole with real rapidity $\lambda_h$, we have \be
\Phi_h(\lambda)=-K(\lambda-\lambda_h).\label{phih}
\ee
In the case of a particle ($\lambda_p$ real or complex), we have\be
\Phi_p(\lambda)=K(\lambda-\lambda_p).\label{phip}
\ee

In the thermodynamic limit, it is convenient to introduce the shorthand notation for the integral operator $\hat{K}$\begin{equation}
\hat{K}\cdot f\left(\lambda\right)\equiv\int_{-B}^{B}d\lambda^\prime\, K\left(\lambda-\lambda^\prime\right)f\left(\lambda^\prime\right),\end{equation}
where $B=\lambda(I_+=N/2)$ is the Fermi boundary. The density of rapidities in the ground state (in the absence of the impurity) is given by the Lieb equation\begin{equation}
\left(1-\frac{\hat{K}}{2\pi}\right)\cdot\rho_{GS}(\lambda) =\frac{p_{0}^{\prime}\left(\lambda\right)}{2\pi}.\label{eq:dressingeq}\end{equation}
The Lieb equation has to be solved consistently with the condition for the average density \begin{equation}
\int_{-B}^{B}d\lambda\,\rho_{GS}(\lambda)=n.\end{equation}

The derivation of the finite size spectrum is standard and is detailed in  Appendix \ref{app:fs}. The result for a single high-energy particle is
\be
\Delta E=\epsilon(\lambda_{p})+\frac{2\pi v}{L}\left[\frac{(\Delta N-n^{p}_{imp})^2}{4K}+K\left(D-d^{p}_{imp}\right)^{2}\right].\label{fsBAparticle}\ee
Here $\epsilon(\lambda)$ is the dressed energy defined by the integral equation\cite{korepin}\be
\left(1-\frac{\hat{K}}{2\pi}\right)\cdot\epsilon(\lambda)  = \epsilon_0(\lambda),\label{dressedenergy}
\ee
and $v$ is the renormalized velocity given by \be
v=\frac{\epsilon^{\prime}(B)}{2\pi\rho_{GS}(B)}.\ee
The phase shifts $n^p_{imp}$ and $d^p_{imp}$ are given by the integrals\bea
n^p_{imp}&=&\int_{-B}^{+B}d\lambda\,\rho_{imp}(\lambda,\lambda_p),\label{nimpBA}\\
2d^p_{imp}&=&\int_{-\infty}^{-B}d\lambda\,\rho_{imp}\left(\lambda,\lambda_p\right)\nonumber\\&&-\int_{B}^{\infty}d\lambda\,\rho_{imp}\left(\lambda,\lambda_p\right),\label{dimpBA}
\eea
where $\rho_{imp}$ is the solution to
\be
\left(1-\frac{\hat{K}}{2\pi}\right)\cdot\rho_{imp}(\lambda,\lambda^\prime)  = \frac{K(\lambda-\lambda^\prime)}{2\pi}.\label{inteqforrhoimp}
\ee

Comparing Eq. (\ref{fsBAparticle}) with Eq. (\ref{spectrumshift}), we conclude that $\varepsilon$ is the dressed energy of the single particle\be
\varepsilon=\epsilon(\lambda_p).\label{energyBAcompare}
\ee
The parameters of the unitary transformation, which set the exponents of the correlation functions, are\be
 \frac{\gamma^p_{R,L}}{\pi}=n^{p}_{imp}\pm2Kd^{p}_{imp}.\label{gammasBA}
 \ee
For a single hole, we find $\varepsilon=-\epsilon(\lambda_h)>0$. The phase shifts for the hole case are\bea
 n^h_{imp}&=&-\int_{-B}^{+B}d\lambda\,\rho_{imp}(\lambda,\lambda_h),\\
2d^h_{imp}&=&-\int_{-\infty}^{-B}d\lambda\,\rho_{imp}\left(\lambda,\lambda_h\right)\nonumber\\&&+\int_{B}^{\infty}d\lambda\,\rho_{imp}\left(\lambda,\lambda_h\right).
 \eea
 
\begin{figure}
\includegraphics[width=0.9\hsize,scale=0.9]{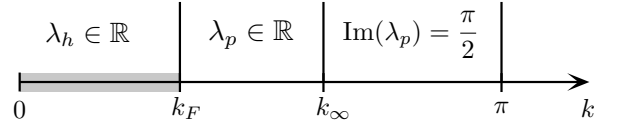}

\caption{Momentum ranges for the three types of excited states considered in the BA calculations for $0<V<2$. For $|k|<k_F$, the simplest excitation is a single hole with real rapidity  below the Fermi boundary, $|\lambda_h|<B$. For $k_F<|k|<k_\infty$, we add a single particle with real rapidity $|\lambda_p|>B$. For $k_\infty<|k|<\pi$, we add a particle with  complex rapidity $\lambda_p=\textrm{ Re}(\lambda_p)+i\pi/2$ (a negative-parity one-string).  \label{fig:kranges}}
\end{figure}

These formulas can be used to compute the exact parameters $\nu_{R,L}$ in Eq. (\ref{nus}) by solving the BA integral equation for $\rho_{imp}$ in Eq. (\ref{inteqforrhoimp}) for a given choice of particle rapidity $\lambda_p$ or hole rapidity $\lambda_h$. In order to analyze the results for  correlation functions as a function of momentum $k$, the latter have to be chosen so that $k(\lambda_{p,h})=k$, where $k(\lambda)$ is the dressed momentum\be
k(\lambda)=p_0(\lambda)-\int_{-B}^{B}d\lambda^\prime \,\theta(\lambda-\lambda^\prime)\rho_{GS}(\lambda^\prime).\label{dressedk}
\ee 
The function $k(\lambda)$ is such that $k(\pm B)=\pm k_F$. Naturally, a single hole excitation  has momentum in the range $0<\left|k(\lambda_h\in\mathbb{R})\right|<k_F$. It follows from Eq. (\ref{dressedk}) that a single-particle excitation with real rapidity has momentum in the range $k_F<|k(\lambda_p\in\mathbb{R})|<k_\infty$, where\be
k_\infty=k_F+\left(\pi-\zeta\right)\left(1-\frac{2k_F}{\pi}\right).
\ee 
A particle excitation with  complex rapidity of the form in Eq. (\ref{lambdastring}) has momentum in the range $k_\infty<\left|k(\lambda_p=\textrm{ Re}(\lambda_p)+i\pi/2)\right|<\pi$. These three types of BA eigenstates cover the Brillouin zone, as illustrated in  Fig. \ref{fig:kranges}. The value of momentum $k_\infty$ is reached by taking $\textrm{ Re}(\lambda_p)\to\infty$. From  Eqs. (\ref{bareenergy}) and (\ref{dressedenergy}), we can see that this is the point where the energy of the single-particle excitation equals the absolute value of the chemical potential\be
\epsilon(k_\infty)=\epsilon_0(\lambda\to\infty)=|\mu|.
\ee
For $0<V<2$, we have $\pi/2<k_\infty<\pi-k_F$.

Since $\varepsilon$ in Eq. (\ref{energyBAcompare}) varies continuously with $\lambda_{p,h}$, the velocity $u$ in Eq. (\ref{Hafterscaling}) is fixed by linearizing the dispersion of the particle excitation around $\lambda_{p,h}$\be
u=\left.\frac{d\epsilon}{d k}\right|_{\lambda_{p,h}}=\frac{\epsilon^\prime(\lambda_{p,h})}{k^\prime(\lambda_{p,h})}.
\ee

In Appendix \ref{app:Ffunction}, we show that $n^{p,h}_{imp}$ and $d^{p,h}_{imp}$ can also be expressed in terms of the shift function as done in Ref. \onlinecite{cheianov}. In the low-energy limit where the momentum of the particle or hole approaches $k_F<\pi/2$, i.e. $\lambda_{p,h}\to B<\infty$,   we use formulas (\ref{nZ1}) and (\ref{dxi1}) in the appendix and obtain\bea
\frac{\gamma^{p,h}_R}{\pi}&\to&\pm(2\sqrt{K}-1-K),\\
\frac{\gamma^{p,h}_L}{\pi}&\to&\pm(K-1).
\eea
This implies that the exponent for the long-time decay of $G_{p,h}(k,t)$  in Eqs. (\ref{Gktdecay})  and (\ref{Ghktdecay}) for $k\to k_F$ is\be
\nu(k\to k_F<\pi/2)=1-\sqrt{K}-\frac{1}{\sqrt{K}}+\frac{1}{2K}+\frac{K}{2}.\label{nukF}
\ee
Thus the exponent of the single-particle singularity of the spectral function in the low-energy limit assumes the value\be
1-\nu(k\to k_F<\pi/2)=\sqrt{K}+\frac{1}{\sqrt{K}}-\frac{1}{2K}-\frac{K}{2}.
\ee
This low-energy exponent, expressed in terms of the Luttinger parameter, is actually universal (see Ref. \onlinecite{imambekov}). On the other hand, we showed in Ref. \onlinecite{pereira} that in the limit of half-filling, $k_F\to\pi/2$, $B\to\infty$ and $|k(\lambda_{p,h})|\neq k_F$, we have\be
n^{p,h}_{imp}\to \pm(K-1)\quad, \quad d^{p,h}_{imp}\to0,\label{nimphalffill}
\ee
independent of particle or hole momentum. In this limit, the region of validity for the particle excitations with real rapidities shrinks to zero as $I_\infty\to N/2=L/4$ and $k_\infty\to k_F=\pi/2$. The single hole and negative parity one-string are still allowed ($\tilde{I}_\infty\to L/4$). As a result, in the limit of half-filling, \be
\left.\nu(k\neq k_F)\right|_{k_F\to\pi/2}\to\frac{K}{2}+\frac{1}{2K}-1\equiv\nu_{ll}.\label{nuhalffill}
\ee

For comparison, the Luttinger liquid result for the particle or hole Green's function (the Luttinger model is particle-hole symmetric due to the linearization of the dispersion) is\cite{luther,meden}\bea
G(x,t)&=&\left[\frac{e^{ik_Fx}}{2\pi(x-vt+i\eta)}-\frac{e^{-ik_Fx}}{2\pi(x+vt-i\eta)}\right]\nonumber\\&&\times\left[\frac{\eta^2}{x^2-(vt-i\eta)^2}\right]^{\nu_{ll}/2}.\label{Gxtluttinger}
\eea
Taking the Fourier transform to momentum space, we get the long-time decay \be
G(k\approx k_F,t)\sim t^{-\nu_{ll}}.
\ee
Note that the exact exponent for the decay of $G_{p,h}\left(k\approx k_F,t\right)$ in Eq. (\ref{nukF}) is \emph{smaller} than the Luttinger liquid exponent $\nu_{ll}$. The Luttinger liquid result for the singularity of the spectral function on the single-particle energy is\cite{meden,voit}\be
A(k,\omega\approx v\delta k)\sim \delta k^{\nu_{ll}/2}(\omega-v\delta k)^{-1+\nu_{ll}/2},\label{Aluttinger}
\ee
where $\delta k\equiv k-k_F$.
Interestingly, according to Eq. (\ref{nuhalffill}), in the limit of half-filling the exact exponent $\nu$ of the long-time decay of $G_{p}(k,t)$ (for $|k|>k_F$) or $G_h(k,t)$ (for $|k|<k_F$) becomes independent of $k$ (for any $k$ in the entire Brillouin zone) and coincides with the Luttinger liquid exponent $\nu_{ll}$ . However, the exact exponent for the single-particle  singularity of $A(k,\omega)$ still differs from the  Luttinger liquid result (note the factor of $1/2$ in Eq. (\ref{Aluttinger}) compared to Eqs. (\ref{Apsingu>v}), (\ref{Apsingu<v}) and (\ref{Ahsingu<v})). The difference stems from the fact that in the Luttinger model the singularity of the spectral function at $\omega\approx v\delta k$ is controlled by the singularity of the Green's function in the vicinity of the light cone $x\approx -vt$, rather than by the exponent for $t\gg |x|$.

That the half-filling exponent can be expressed solely in terms of the Luttinger parameter is a peculiarity of the integrable model. The exception is the model with  $V=2$, which is equivalent to the Heisenberg spin chain. If additional interactions which break integrability still respect SU(2) symmetry, the exponent $\nu=\nu_{ll}=1/4$ for $K=1/2$ is universal.\cite{imamb3}

\subsection{One-two-string one-hole excitations\label{sec:BAbound}}
The Bethe equations in Eq. (\ref{BAequation}) admit solutions with  a pair of complex rapidities of the form\be
\lambda_s=w+ i\zeta/2,\quad \lambda_s^*=w-i\zeta/2,\quad w\in\mathbb{R}.
\ee
An eigenstate with two particles described by such pair of rapidities is called a two-string.\cite{sutherland} The fact that $e^{i\theta(\lambda_s-\lambda_s^*)}=0$ guarantees that the wave function of the two-string is normalizable and describes a two-particle bound state. 

From Eqs. (\ref{p0}) and (\ref{bareenergy}), we have that the bare momentum and energy of the two-string are\be
\mathcal{P}_0(w)=p_0(\lambda_s)+p_0(\lambda_s^*)=i\log\left[-\frac{\sinh(i\zeta+w)}{\sinh(i\zeta-w)}\right].\ee
\be
\mathcal{E}_0(w)=\epsilon_0(\lambda_s)+\epsilon_0(\lambda_s^*)=-\frac{2\cos\zeta\sin^2\zeta}{\cosh^2w-\cos^2\zeta}-2\mu.
\ee
Note that the bare momentum of the two-string is restricted to the interval\be
|\mathcal{P}_0(w) -\pi|\textrm{ mod } 2\pi<\pi-2\zeta.
\ee

If the two-string is added to the ground state of $N$ fermions, the renormalized momentum and energy must include the backflow of the ground state. In analogy with Eqs. (\ref{dressedenergy}) and (\ref{dressedk}), the dressed momentum and energy of the two-string excitation are\bea
\mathcal{P}(w)&=&\mathcal{P}_0(w)-\int_{-B}^{+B}d\lambda^\prime\,\Theta(w-\lambda^\prime)\rho(\lambda^\prime),\\
\mathcal{E}(w)&=&\mathcal{E}_0(w)+\int_{-B}^{+B}\frac{d\lambda^\prime}{2\pi}\,\mathcal{K}(w-\lambda^\prime)\epsilon(\lambda^\prime),
\eea
where\bea
\Theta(w)&=&\theta(w+i\zeta/2)+\theta(w-i\zeta/2),\\
\mathcal{K}(w)&=&K(w+i\zeta/2)+K(w-i\zeta/2).
\eea
One can then verify that the dressed momentum is restricted to \be
|\mathcal{P}(w)-\pi| \textrm{ mod } 2\pi <\Lambda_{bs},
\ee
where \be
\Lambda_{bs}\equiv \pi-\mathcal{P}(-\infty)=(\pi-2\zeta)\left(1-\frac{2k_F}{\pi}\right).
\ee
This fixes the momentum range for the bound state band alluded to in Sec. \ref{sec:bound}. The minimum value of momentum $k$ for which a one-two-string one-hole excitation exists is\be
k_b=k_F+2\zeta\left(1-\frac{2k_F}{\pi}\right).
\ee
Moreover, it is easy to show that the energy of the bound state at the edge of the bound state band is\be
\mathcal{E}(\mathcal{P}=\pi\pm\Lambda_{bs})=\mathcal{E}(w\to\pm\infty)=2|\mu|.
\ee
In the limit of half-filling, $k_F\to\pi/2$, the bound state band shrinks to zero ($\Lambda_{bs}\to0$) for arbitrary values of the interaction strength and the region of validity of the power-law singularity vanishes.

The phase shifts at the Fermi boundaries due to the creation of a two-string can be calculated by the methods described in the previous section. Defining $\varrho_{imp}(\lambda,u)$ by the integral equation\be
\left(1-\frac{\hat{K}}{2\pi}\right)\cdot \varrho_{imp}(\lambda,w)=\frac{\mathcal{K}(\lambda-w)}{2\pi},
\ee
we can write the phase shifts that enter the finite size spectrum as\bea
n_{imp}^{bs}&=&\int_{-B}^{+B}d\lambda\,\varrho_{imp}(\lambda,w_{0}),\\
2d_{imp}^{bs}&=&\int_{-\infty}^{-B}d\lambda\,\varrho_{imp}(\lambda,w_{0})\nonumber\\
&&-\int_{B}^{\infty}d\lambda\,\varrho_{imp}(\lambda,w_{0}),
\eea
where $w_0$ is chosen such that $\mathcal{P}(w_0)=P_0$ for a bound state with center-of-mass momentum $P_0$. By correspondence with the finite size spectrum of the field theory, the parameters of the unitary transformation that determine the exponents in Eq. (\ref{nusbound}) are given by\be
\frac{\gamma^{bs}_{R,L}}{\pi}=n_{imp}^{bs}\pm 2Kd_{imp}^{bs}.
\ee

The velocity of the bound state is obtained by linearizing the dispersion about  $w_0$\be
u_{bs}=\frac{\mathcal{E}^\prime(w_0)}{\mathcal{P}^\prime(w_0)}.
\ee

\section{Density Matrix Renormalization Group Results for fermion Green's function\label{sec:DMRG}}
\subsection{Method}
We used real time DMRG methods (tDMRG) to calculate the spectral functions for this system.
The typical size of the systems studied was $L=400$. The main details of the techniques used
were described briefly in Ref. \onlinecite{pereira}, and in considerable detail
in Ref. \onlinecite{spin1}. Here we summarize the method, and also discuss
a finite size issue which did not come up in the earlier work, and how we treat it.

After finding the ground state with
ground state DMRG, an operator $\psi^{\phantom\dagger}$ or $\psi^\dagger$ was applied to a site in the
center, and the resulting solution to the time-dependent Schr\"odinger equation was used to
calculate a space-time dependent hole Green's function \be G_h(x,t)=\bra\psi^\dagger_{j+x}(t)\psi^{\phantom\dagger}_j(0)\ket\ee
 or particle Green's function\be
 G_p(x,t)=\bra\psi^{\phantom\dagger}_{j+x}(t)\psi^\dagger_j(0)\ket\ee  
for $t>0$ out to a 
particular time $T \sim 20-30$.  The hole and particle Green's functions 
each require a separate simulation.
The result was Fourier transformed in space.  For each desired wave vector $k$, the resulting
time-dependent functions $G_{p,h}(k,t)$ as defined in Eqs. (\ref{Gpktparticle}) and (\ref{Ghktdecay}) 
were either fit to an asymptotic long-time form and extrapolated
to very long times using it, or extrapolated to long times using
a very general method called linear prediction.  In both cases,
a fast Fourier transform was used to obtain high resolution
spectra.  The asymptotic form also gave directly exponents characterizing the frequency
singularities.  Typically, we specified a truncation error of $10^{-8}$ to set the 
number of states kept per block $m$, with an additional constraint $m \le 1500$.  These
parameters were varied to determine errors in $G_{p,h}(x,t)$, and the typical errors for the
parameters used were $10^{-4} - 10^{-5}$.  The functions $G_{p,h}(x,t=0)$ decay as a power law
in $x$, potentially giving large finite size effects.  However, these tails in $x$ only appear
in the real part of $G_{p,h}$, and it is easy to reconstruct the entire spectrum from only the
imaginary part when the particle and hole Green's functions are treated separately.  
The support of the imaginary part spreads out from the center site with a speed given
by the maximum velocity $v_m$ as a function of $k$ of the hole or particle; finite size
effects are small as long as $v_m T \ll L/2$.
The long tails in $x$ are associated with $\omega\approx 0$; the determination
of spectra using only the imaginary part generates an odd function in frequency, eliminating
the $\omega=0$ contribution.

\begin{figure}

\includegraphics*[width=0.95\hsize,scale=1.0]{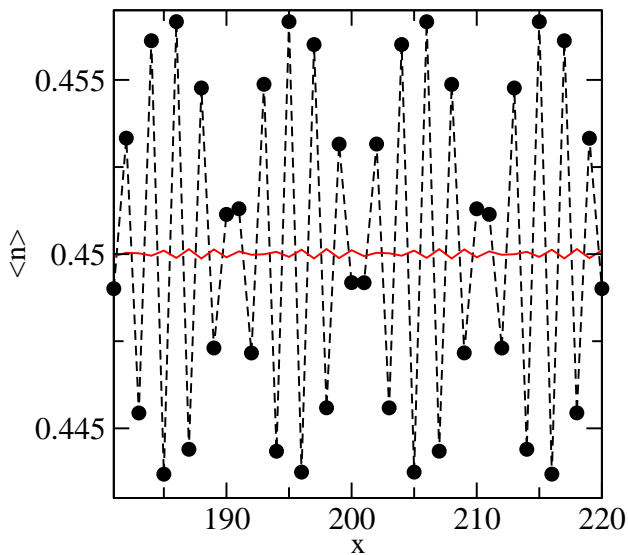}

\caption{(Color online) Local density $\bra n(x)\ket$ in the ground state for an average density of $n=0.45$ and interaction strength $V=1$, near the the center of a system with system size $L=400$. Black dots: open boundary conditions; solid red line: smooth boundary conditions. \label{fig:friedel}}
\end{figure}

\begin{figure}

\includegraphics*[width=0.9\hsize,scale=1.0]{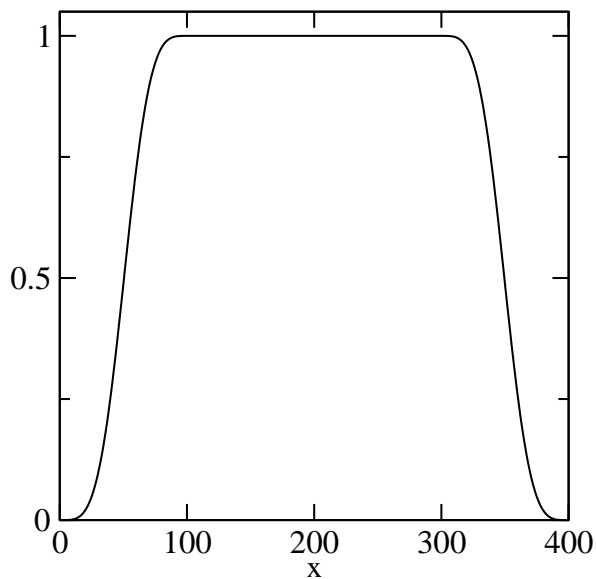}

\caption{Smoothing function $s(x)$ used in the DMRG runs with smooth boundary conditions. \label{fig:smoothfunc}}
\end{figure}

Another finite size effect comes from the Friedel oscillations induced by the boundaries of
an open system in the ground state. 
In Fig. \ref{fig:friedel}, the solid dots represent the ground state density on each
site $\langle n(x) \rangle$ of an open $L=400$ system with a desired average density of 0.45.
The open boundaries induce oscillations in the density which decay slowly away from the ends.
The density is directly related to $G_{h}(x=0,t=0)$, and these oscillations introduce a finite
size error in this system of order 1\%.  There are several ways one could substantially
reduce the finite size effects.  The simplest (conceptually) would be to use periodic boundary conditions,
which can now be treated within DMRG with only a slight computational penalty.\cite{pippan}
Because our most efficient computer program for real time evolution does not incorporate the
new periodic algorithm, we have chosen to use smooth boundary conditions (BCs).\cite{vekic}
For smooth BCs,
one introduces a chemical potential to set the density in the center of the system, but one
still works with a fixed number of particles. The parameters of the Hamiltonian and the chemical
potential are held fixed at bulk values in a central region of the system, but near the edges
they are scaled down to zero smoothly; see Fig. \ref{fig:smoothfunc}.  The basic smoothing function
by which the parameters were multiplied  (after scaling and shifting)
was \be
s(x) = \frac{1}{2}\left\{1 + \sin\left[\frac{\pi}{2} \cos(\pi x)\right]\right\},\ee which falls smoothly from 1 to 0
over $0 < x < 1$, and has three vanishing derivatives at the endpoints. The smoothing regions
were one fourth the lattice length at each end. The smooth boundaries allow the central bulk region to
minimize its energy, largely eliminating the Friedel oscillations,
at the expense of large oscillations near the low-energy edges.  The chemical potential allows
one to fix the bulk density to any value, even irrational. One sets the overall number of particles
to the integer giving the closest overall density, and the system adjusts the density at the edges
to give a density in the center determined by the chemical potential.  (This means that a number
of ground-state-only runs must be done to determine the correct chemical potential for the given
density; these contribute little to the overall computational cost.)
The smooth connection between
the boundary region and the bulk is essential to avoid artifacts there.  Fig. \ref{fig:friedel}
shows the resulting ground state density, with acceptably small oscillations in the central region.
The real time evolution is performed using the smoothed Hamiltonian.  The evolution should be
stopped before the signal initiated by $\psi$ or $\psi^\dagger$ in the center reaches the edge
of the bulk region.
We have checked how well
the smooth BCs perform in determining the correct thermodynamic limit Green's function in the
case of $V=0$, where one can obtain numerically exact results from diagonalizing and manipulating
$L\times L$ matrices.  We find finite size errors of $10^{-3} - 10^{-4}$ for the size systems we use.

The smooth BCs duplicate the behavior of a very large system in another respect:  the entanglement
entropy, which governs the number of states kept per block $m$ for a given error, is significantly
larger than with open BCs for a given $L$. We have not studied this interesting effect in detail. We may obtain a simple understanding of this effect by considering two weakly coupled spins at the two ends of a Heisenberg antiferromagnetic spin chain of even length.  If these end spins are sufficiently weakly coupled then they will form a singlet together in the ground state, whereas all other spins go into a complicated ``resonating valence bond" state. This long range singlet increases the entanglement of the left and
right halves of the system.  Related entanglement phenomena in the case of a
single weakly coupled spin were studied in Ref.  \onlinecite{sorensen}. We were able to increase the number of states kept sufficiently to obtain accurate spectral functions
despite the larger entanglement. An intuitive picture explaining this effect is that the smooth
wave function might be written as a superposition of a non-translationally invariant wave function,
such as shown by the circles in Fig. \ref{fig:friedel}, with translations of itself, sufficient
to eliminate the static oscillations. If this idea is roughly right, then one might obtain accurate
spectra by using one open BC lattice, but averaging over the starting point where $\psi$ or $\psi^\dagger$
is initially applied, say with a Gaussian envelope of width $5-10$, 
to smooth out the effect of the oscillations.
We have tried this approach as another check on the smooth BC results.  It requires at least 5-10 
runs, one
for each starting point, but each has low entanglement. The results were quite satisfactory,
giving good agreement with the smooth BC results.

\subsection{Results}
\begin{figure}

\includegraphics*[width=0.9\hsize,scale=1.0]{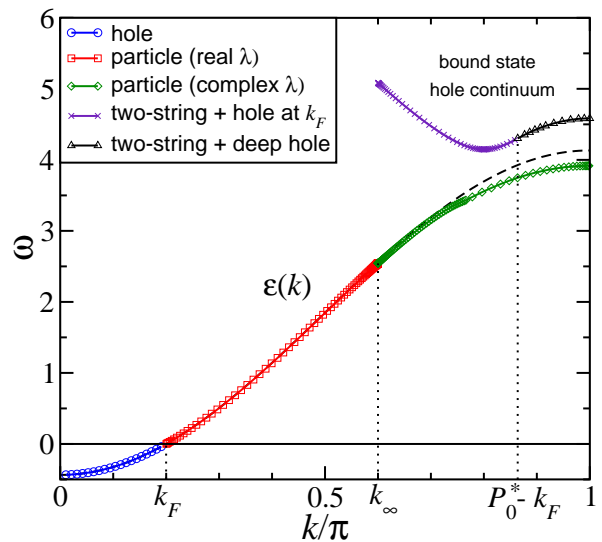}

\caption{(Color online) Exact single particle dispersion and lower threshold of the one-two-string one-hole continuum for interaction strength $V=1$ and fermion density $n=0.2$. The curve $\omega=\epsilon(k)$ is divided into the hole region ($k<k_F$) and two particle regions (particle excitations with real rapidities for $k_F<k<k_\infty$ and complex rapidities for $k_\infty<k<\pi$). The dashed line is a cosine function fitted to the hole region of the spectrum, showing that the renormalized dispersion deviates from the $\cos k$ dependence. The  one-two-string one-hole continuum is defined for $k_b<k<\pi$ (see  text). For $k_b<k<P_0^*-k_F$, the lower threshold is defined by a hole at $k_F$ and a two-string with momentum $k+k_F$. For $P_0^*-k_F<k<\pi$, the lower threshold is defined by a deep hole and a two-string with the same velocity. \label{fig:dispersion}}
\end{figure}

We used the BA integral equations in Sec. \ref{sec:BA} to  numerically evaluate the dressed energies $\epsilon(\lambda)$ and $\mathcal{E}(w)$ of the elementary excitations for the model with $V=1$ and $n=0.2$. The values of some important parameters are presented in Table \ref{tab:param}.  Note that for $V=1$ we have $k_b=k_\infty$, thus the range of momentum where the bound state contribution exists coincides with the range where the single-particle excitation is described by a negative-parity one-string. The result for  the dispersion of the single-particle excitations and the lower threshold of the one-two-string one-hole continuum is shown in Fig. \ref{fig:dispersion}. We find that, for $n=0.2$ and $V=1$, the velocity of the two-string equals the renormalized Fermi velocity at $P_0^*\approx3.342<\pi+\Lambda_{bs}$. As a result, the nature of the bound state singularity changes at $k=P_0^*-k_F\approx2.714$ (see Fig. \ref{fig:dispersion}).

\psfrag{kinf}{$k_\infty$}
\psfrag{pi-kF}{$\pi-k_F$}
\psfrag{kF}{$k_F$}
\psfrag{P0*-kF}{$P_0^*-k_F$}

\begin{table}[t]
\caption{Parameters for the model with $V=1$ and $n=0.2$, obtained either from formulas given in Sec. \ref{sec:BA} or from the numerical solution of the BA equations for finite Fermi boundary $B$.\label{tab:param}}
\begin{tabular}{ccccccc}
\hline
$h$&
$v$&
$K$&
$k_F$&
$k_\infty$&
$k_b$&
$P_0^*-k_F$
 \\
\hline 
-2.542&
1.343&
0.873&
0.628&
1.885&
1.885&
2.714\\
\hline

\end{tabular}
\end{table}

\begin{figure}
\includegraphics*[width=0.95\hsize,scale=1.0]{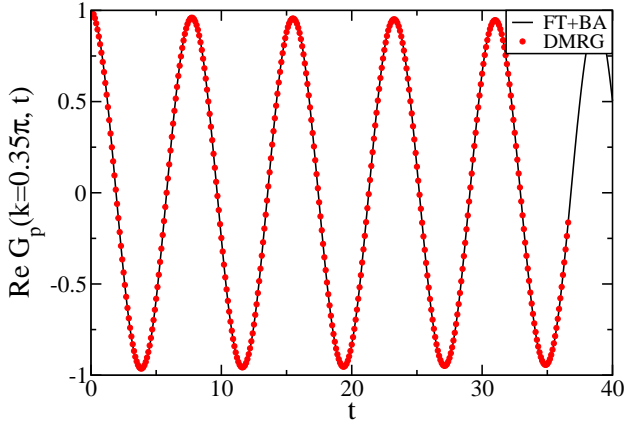}

\caption{(Color online) Real part of the particle Green's function $G_p(k=0.35\pi,t)$ for  $V=1$ and  $n=0.2$. Red dots are numerical tDMRG results and solid line (FT+BA) is the field theory result  of Eq. (\ref{Gktdecay}) with parameters computed from the Bethe ansatz: frequency  $\varepsilon^{BA}=0.8105$, exponent $\nu^{BA}=0.0097$ and phase $\nu^{BA}_L-\nu^{BA}_R=0.0076$. Only the amplitude has been fitted. If we fit the DMRG results for $t>5$ with four free parameters, we find the best fit for $\varepsilon^{DMRG}=0.8104$, $\nu^{DMRG}=0.0109$ and  $\nu^{DMRG}_L-\nu^{DMRG}_R=0.0074$. \label{fig:Gkt204}}
\end{figure}

Using the tDMRG method, we calculated both particle and hole Green's function for the values  $n=0.2$ and $V=1$ and for times out to $T=37$. The tDMRG results show that the particle Green's function $G_p(k,t)$ becomes $O(1)$ for $|k|>k_F$. Due to interactions, $G_p(k,t)$ is nonzero for momenta below the Fermi surface, but its amplitude is much smaller. Similarly, the hole Green's function is $O(1)$ for $|k|<k_F$.  We restrict our analysis to the ranges $|k|>k_F$ for the particle Green's function and $|k|<k_F$ for the hole Green's function. As an example, Fig. \ref{fig:Gkt204} shows the tDMRG result for the particle Green's function $G_p(k,t)$ for $k=0.35\pi$. This is in the fast particle regime $u>v$. The numerical result is compared with  the field theory result  of Eq. (\ref{Gktdecay}), which predicts  an asymptotic behavior with oscillations at a single frequency and  pure power-law decay. The BA fixes the frequency, phase and  exponent of the decay of $G_p(k,t)$.  Such very good agreement  of Eq. (\ref{Gktdecay})  with the tDMRG result is typical for $k$ values in the regime $u>v$. This lends support to  the conjecture that the decay rate $1/\tau$ discussed in Sec. \ref{sec:decaysmallk} is exactly zero for the integrable model. 

Similar results with a pure power-law decay are obtained for the hole Green's function $G_h(k,t)$. In this case we expect $1/\tau$ to vanish due to kinematic constraints, since the single hole is at the threshold of the multiparticle continuum. (The exact edges of the support can be determined from the dispersion in Fig. \ref{fig:dispersion} and  are similar to the curves in Fig. \ref{fig:support}a.)

\begin{figure}
\includegraphics*[width=0.95\hsize,scale=1.0]{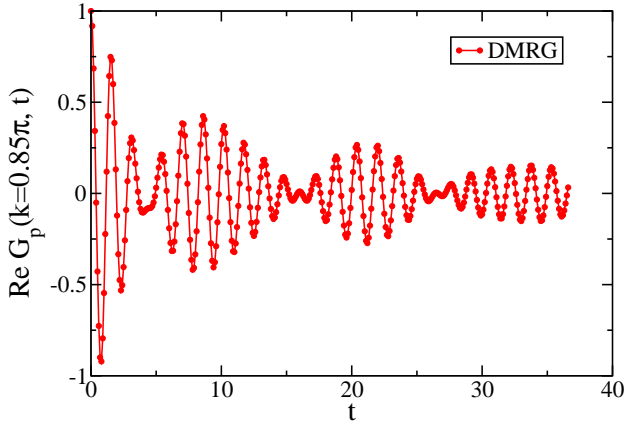}

\caption{(Color online) tDMRG result for the real part of the particle Green's function for $V=1$, $n=0.2$ and momentum $k=0.85\pi$. The line is a guide to the eye.  \label{fig:Akt267}}
\end{figure}

For contrast, we show in Fig. \ref{fig:Akt267} the tDMRG results for the particle Green's function for $k=0.85\pi$. The data is typical of that for  values of $k$ near $\pi$ and shows the presence of two dominant frequencies in the long-time behavior.

\begin{figure}
\includegraphics*[width=0.95\hsize,scale=1.0]{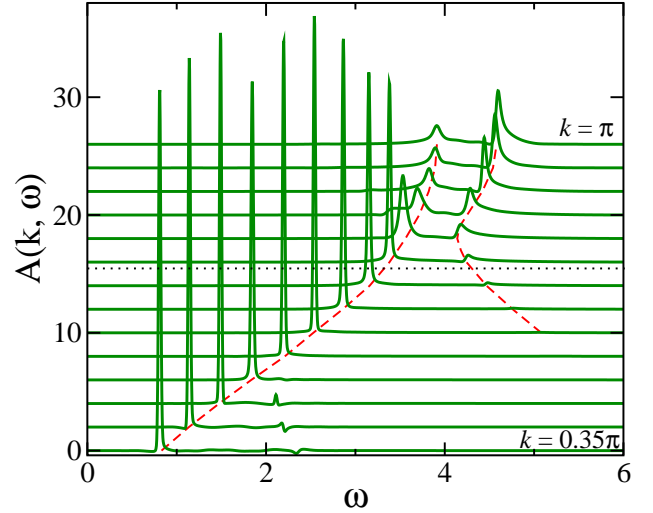}

\caption{(Color online) Particle spectral function $A_p(k,\omega)$ for $V=1$, $n=0.2$ and  $k/\pi=0.35, 0.4, 0.45, ..., 1$, obtained by Fourier transforming the tDMRG data using the linear extrapolation method. The curves for different values of $k$ are shifted vertically ($k$ increases from bottom to top). The horizontal dotted line ($k\approx 2.31$) separates the regime $u>v$ from the regime $u<v$. For $k_F<k<k_b$, there is only one peak at the energy of the single-particle excitation. For $k_b<k<\pi$, there appears a second peak, which we interpret as due to a two-particle bound state and a free hole. The energy of the second peak is non-monotonic in $k$ and has a  minimum at $k=\pi-k_F$. The dashed lines are the BA predictions for the dispersion relations in Fig. \ref{fig:dispersion}.  \label{fig:Akwpeaks}}
\end{figure}

By utilizing the linear extrapolation method described in Ref. \onlinecite{spin1}, we computed the Fourier transform of the real time tDMRG data to produce line shapes for the spectral function without any analytical input. We confirm the  simple behavior of the hole spectral function, with a single sharp peak on the single-hole energy. We thus focus on the particle part of the spectrum, which shows a more interesting behavior.   The results for $k\geq0.35\pi$ are shown in Fig. \ref{fig:Akwpeaks}.  The energies of the peaks in the tDMRG results agree with the energies predicted by the BA solution shown in Fig. \ref{fig:dispersion}. The single-particle peaks remain sharp for fairly large values of $k-k_F$  which are in the regime $u>v$ ($k<2.31$ according to the BA dispersion in Fig. \ref{fig:dispersion}).    On the other hand, the broadening of the single-particle peak is apparent for the larger  values of $k$, near $k=\pi$. This happens in the slow particle regime $u<v$, where we expect a nonzero decay rate due to the process discussed in Sec. \ref{sec:decaylargek}.

The second peak associated with the one-bound-state one-hole excitation shows up for momentum $k>k_b$, as predicted by the BA. For $k$ near $k_b$, the second peak is small and broad and hardly visible in Fig. \ref{fig:Akwpeaks}. This may be explained qualitatively by arguing that, when the velocity of the two-particle bound state is smaller than the Fermi velocity, the decay of the bound state by scattering of one low-energy particle-hole pair  is kinematically possible (for reasons analogous to the arguments given in Sec. \ref{sec:decaylargek}), so the edge singularity may  actually be broadened.

As $k$ increases towards $k=\pi$, spectral weight is transferred from the single-particle peak to the bound-state hole continuum. The second peak also becomes sharper with increasing $k$, in accord with a stronger power-law singularity in the regime where the lower threshold is defined by a bound state and a deep hole with the same velocity (see Sec. \ref{sec:bound}).

\begin{figure}
\includegraphics*[width=0.95\hsize,scale=1.0]{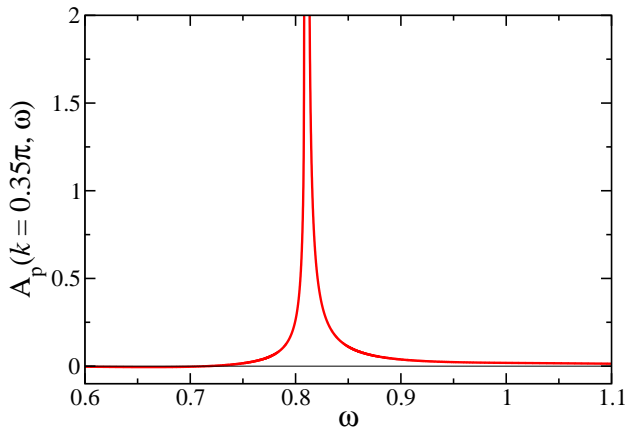}

\caption{(Color online) Spectral function in the vicinity of the single-particle energy for $V=1$, $n=0.2$ and $k=0.35\pi$, obtained by extending the tDMRG data shown in Fig. \ref{fig:Gkt204} to long times using the field theory formulas and then taking the Fourier transform.  \label{fig:Akw204}}
\end{figure}

\begin{figure}
\includegraphics*[width=0.95\hsize,scale=1.0]{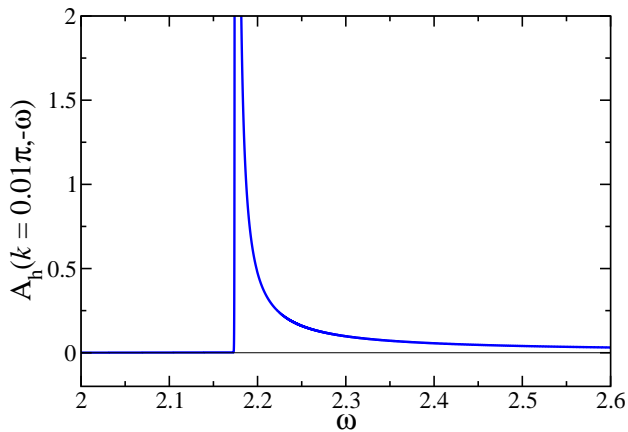}

\caption{(Color online) Spectral function in the vicinity of the single-hole energy for $V=1$, $n=0.45$ and $k=0.01\pi$, obtained by extending the tDMRG data as in Fig. \ref{fig:Akw204}.  \label{fig:Akw000}}
\end{figure}

We obtain high resolution spectral functions by using the asymptotic behavior predicted by the field theory to extend the tDMRG data to arbitrarily large $t$ before computing the Fourier transform. The result of this combination of methods is illustrated in Fig. \ref{fig:Akw204} for the particle spectral function in the fast particle regime, $u>v$. In Fig. \ref{fig:Akw000} we show the  hole spectral function for a different value of density, $n=0.45$. (For $n=0.2$, the maximum energy of a hole is small, so the tDMRG data with $t<40$ is not representative of the long time behavior.)    While the spectral function  exhibits a two-sided power-law singularity on the single-particle energy, it vanishes below the single-hole energy. These line shapes represent a high-energy extrapolation of the low-energy result of Khodas et al.,  \cite{khodas} except for the absence of broadening of the single-particle peak. We are not able to access directly the regime $k\approx k_F$ because in the low energy limit the region of validity of the power law in Eq. (\ref{Apsingu>v})  becomes extremely small. This means that it would be necessary to go to extremely large values of $t$ in the tDMRG calculations in order to observe the asymptotic behavior of $G_p(k,t)$. The contribution of the convergent edge singularities to the long-time decay of $G_p(k,t)$ is also rather small, so we have not attempted to extract the behavior near the multiparticle thresholds where $A(k,\omega)$ vanishes.

We have also calculated the particle Green's function for the same $V=1$ and $n=0.45$, which is closer to half-filling. In this case both hole and particle spectral function show a single sharp peak at $\omega=\epsilon(k)$. We have not observed a second peak in the particle spectral function for $k$ near $\pi$. This is in agreement with the prediction that the contribution of the one-bound-state one-hole excitation disappears as we approach half-filling (see Sec. \ref{sec:BAbound}).

\begin{figure}
\includegraphics*[width=0.9\hsize,scale=1.0]{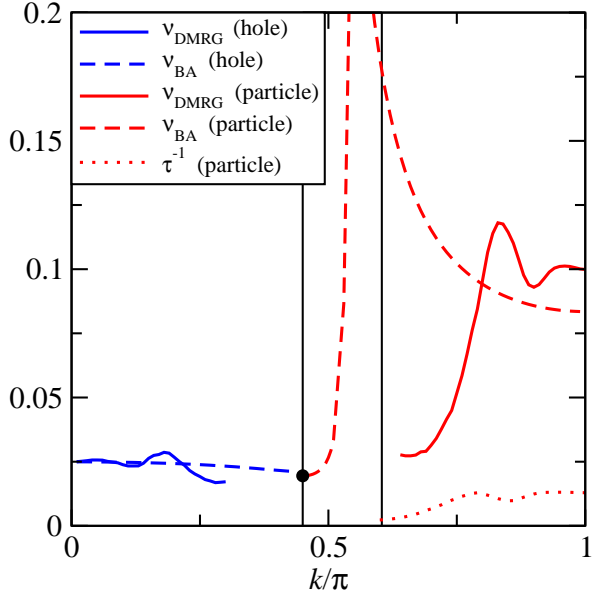}

\caption{(Color online) Exponent of particle (for $k>k_F$) or hole (for $k<k_F$) Green's function for $n=0.45$ and $V=1$. The solid lines represent the exponents extracted from tDMRG by fitting the data in the  time interval $10<t<20$ to Eq. (\ref{Gexpdecay}). The dotted line is the decay rate of the single particle extracted from the fitting. The dashed lines represent the exponent calculated numerically from the BA.  The dot at $k=k_F$ indicates the universal low-energy exponent in Eq. (\ref{nukF}). The vertical lines indicate the values of momentum $k=k_F=0.45\pi$ and $k\approx1.897$, near which the tDMRG exponents  are least reliable.\label{fig:expDMRGvsBA}}
\end{figure}

Assuming a single frequency is able to describe the long-time decay of particle and hole Green's function, we  fitted the data to Eq. (\ref{Gexpdecay}) in order to extract a possible nonzero decay rate. Eq. (\ref{Gexpdecay}) is valid  for $k$ values in the regime $u<v$, in which a single parameter fixes both phase and exponent.  The results are shown in Fig. \ref{fig:expDMRGvsBA}. We first note that fitting  the tDMRG data up to a maximum time $T=20$ is not reliable near two points. The first point is $k=k_F$, where the frequency of the time oscillation vanishes. The second point is $k\approx1.897$, where $u\approx v$ according to the BA. Near this point, the time scale for reaching the asymptotic decay also diverges (see Eq. (\ref{Gktpart})). These two points are indicated by vertical lines in Fig. \ref{fig:expDMRGvsBA}. The agreement between the tDMRG and BA exponents is expected to be best near $k=0$ for the hole case and near $k=\pi$ for the particle case. 

For the hole Green's function, we find  very good agreement between the numerical  exponent and the one predicted by the BA. Moreover, in this case the numerical decay rate is negligible ($<10^{-4}$ for all values of $k<k_F$). For the particle Green's function, the agreement between BA and tDMRG exponents is not as good as for the hole case. This is expected given that we are fitting with two small parameters, the power law decay exponent and the decay rate. We have checked that fitting the tDMRG results with the exponents constrained by the BA works as well as fitting with unconstrained exponents. Thus the oscillations in the tDMRG exponents in Fig. \ref{fig:expDMRGvsBA} are believed to be due to the error in the fitting. Nonetheless, the tDMRG results clearly show that  the decay rate  $1/\tau$ is nonzero for a high-energy particle in the regime $u<v$. The results also suggest that $1/\tau$ decreases as $k$ approaches the value where $u= v$, possibly vanishing at the lower threshold.  We note that the behavior of the BA exponent is qualitatively similar to the  expected from the weak coupling expressions  in Eq. (\ref{gammau-v}).

Finally, we point out that the long-time behavior of the fermion Green's function in real space is dominated by a saddle point contribution.\cite{pereira} The latter corresponds to a hole at the bottom of the band, $k=0$, in the case of the hole Green's function, or to a particle at the top of the band, $k=\pi$, in the case of the particle Green's function. The high-energy  excitations near these points have parabolic dispersion $\epsilon(k)\approx\varepsilon+k^2/2m$, where $m<0$ is the renormalized mass. Thus the propagator for the decoupled $\bar{d}$ particle in Eq. (\ref{propfreepart})  reads\be
G^{(0)}_p(x,t)\sim \sqrt{\frac{m}{2\pi it}}e^{-i\varepsilon_\pi t-\frac{imx^2}{2t}}.
\ee
Therefore, in the noninteracting case, this contribution to the particle Green's function  oscillates at a high frequency $\varepsilon_\pi\sim O(1)$ for $t\gg |m|x^2$ and decays as $1/\sqrt{t}$. In the interacting case, the frequency of the oscillation becomes the renormalized energy of the single particle with momentum $k=\pi$.  In addition, the exponent is modified  by the coupling to the low-energy modes.   In the case of the particle Green's function, there is also an exponential decay associated with the decay rate $1/\tau_\pi$ of the particle at $k=\pi$ when the model is below half-filling. The long-time behavior of the particle Green's function is then given by\be
G_p(x,t\gg x/v,|m|x^2)\sim \frac{Ae^{-i\varepsilon_\pi t-t/\tau_\pi}}{t^{\eta_\pi}}+\frac{B}{t^{1+\nu_{ll}}}.\label{topoftheband}
\ee
Here $A$ and $B$ are complex amplitudes. The second term in Eq. (\ref{topoftheband}) is the standard Luttinger liquid result, which does not oscillate in time. The renormalized exponent of the high-energy term is \be \eta_\pi=1/2+\nu(k=\pi),\ee
with $\nu(k)$ given by Eqs. (\ref{nus}) and (\ref{nunuRnuL}). This is smaller than the exponent of the Luttinger liquid term; however, for $t>\tau_\pi$ the high-energy term in $G_p(x,t)$ decays exponentially. We note that both a particle at $k=\pi$ and a hole at $k=0$ couple symmetrically to the low-energy modes at the two Fermi points, hence $\nu_R(k=0,\pi)=\nu_L(k=0,\pi)$. Below half-filling the hole Green's function decays as\be
G_h(x,t\gg x/v,|m|x^2)\sim  \frac{A^\prime e^{-i\varepsilon_0 t}}{t^{\eta_0}}+\frac{B^\prime}{t^{1+\nu_{ll}}},
\ee
with $\eta_0=1/2+\nu(k=0)$ and no exponential decay of the high-energy contribution. 

At half-filling, particle and hole Green's functions are equivalent. In this case, we have analytical expressions for the frequencies and exponents\cite{pereira}\bea
\varepsilon_0&=&\varepsilon_\pi=\frac{\pi\sqrt{1-(V/2)^2}}{\arccos(V/2)},\label{halffillenergy}\\
\eta_0&=&\eta_\pi=1/2+\nu_{ll},\label{halffillexp}
\eea
where $\nu_{ll}=(K+K^{-1}-2)/2$ with the Luttinger parameter\cite{giamarchi}\be
K=\frac{\pi}{2[\pi-\arccos(V/2)]}.
\ee
Moreover, the decay rate $1/\tau_\pi$ vanishes at half-filling. 

We have calculated the fermion Green's function $G(x=0,t)$ at half-filling and for various values of interaction strength $V$ using tDMRG. We used simulated annealing to fit the numerical results to the formula\be
G(x,t)\sim  \frac{A e^{-i\varepsilon_{DMRG} t}}{t^{\eta_{DMRG}}}+\frac{B}{t^{\alpha_{DMRG}}}.\label{fitformulaself}
\ee
The results are shown in Table \ref{tableI}. We find good agreement with the analytical BA predictions of Eqs. (\ref{halffillenergy}) and (\ref{halffillexp}) for all values of $V$.

\begin{table}
\caption{Exponents and frequencies for the fermion Green's function $G(x=0,t)$ at half-filling and for $V=0.25,0.5,0.75,1,1.5$. The  parameters $\varepsilon_{DMRG}$, $\eta_{DMRG}$ and $\alpha_{DMRG}$ were obtained by fitting the DMRG results to Eq. (\ref{fitformulaself}). The other parameters are analytical BA predictions and should be compared with the adjacent DMRG results.  \label{tableI}}

\begin{tabular}{c|c|cc|cc|cc}
\hline\hline 
$V$&
$K$&
$\varepsilon_{DMRG}$&
$\varepsilon_0$&
$\eta_{DMRG}$&
$\eta_0$&
$\alpha_{DMRG}$&
$1+\nu_{ll}$\\
\hline 
0&
1&
$-$&
2&
$-$&
0.5&
$-$&
1\\
0.25&
0.926&
2.156&
2.156&
0.503&
0.503&
0.985&
1.003\\
0.5\phantom{0}&
0.861&
2.308&
2.308&
0.510&
0.511&
0.986&
1.011\\
0.75&
0.803&
2.454&
2.454&
0.523&
0.524&
1.002&
1.024\\
1\phantom{00}&
0.75\phantom{0}&
2.598&
2.598&
0.539&
0.542&
1.013&
1.042\\
1.5\phantom{0}&
0.649&
2.876&
2.876&
0.583&
0.595&
1.048&
1.095\\
\hline\hline
\end{tabular}

\end{table}

\section{Dynamical structure factor\label{sec:DSF}}
The methods used in this work can be generally applied to the study of singularities of dynamical correlation functions. The basic idea is to identify the excitations which define the thresholds of the spectrum in the noninteracting limit and then ask how interactions between the particles in the final state affect the behavior of the dynamical function near the threshold. This can be answered with the help of effective models such as Eqs. (\ref{Hafterscaling}). One quantity of interest is the dynamical structure factor\be
S(q,\omega)=\frac{1}{L}\sum_{j,j^\prime}e^{-iq(j-j^\prime)}\int_{-\infty}^\infty dt\,e^{i\omega t} \bra n_j(t)n_{j^\prime}(0)\ket.
\ee
The dynamical structure factor for the Galilean invariant model was studied in Ref. \onlinecite{pustilnik}. In the lattice model, $S(q,\omega)$ is equivalent to the longitudinal dynamical spin structure factor of spin-1/2 chains.\cite{pereira} A particular interesting case from the point of view of experiments is the model with $V=2$, which is equivalent to the Heisenberg spin chain\be
H=J\sum_{j=1}^{L}\vec{S}_j\cdot\vec{S}_{j+1}.\label{heisenbergspinchain}
\ee
In this section we discuss qualitatively the implications of our results on the spectral function for  the line shape of  $S(q,\omega)$ away from half-filling. 

\begin{figure}
\includegraphics*[width=0.95\hsize,scale=1.0]{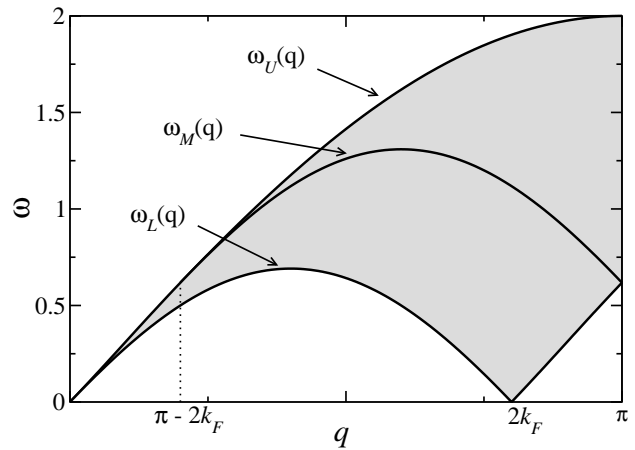}

\caption{Particle-hole continuum for the noninteracting model with $n=0.4$.  The dynamical structure factor $S(q,\omega)$ is nonzero in the shaded area.  \label{fig:2pcont}}
\end{figure}

The excitations which give the dominant contribution to the dynamical structure factor were identified by M\"uller et al.\cite{muller1,muller2} In the noninteracting case, the region where $S(q,\omega)$ is nonzero is given by the particle-hole continuum shown in Fig. \ref{fig:2pcont} for $n=0.4$. For small momentum, $q\ll\pi-2k_F$, $S(q,\omega)$ is  similar to the result for the Galilean invariant model.\cite{pereiraJSTAT}  Instead of discussing all possible cases, here we focus on the momentum range $\pi-2k_F<q<2k_F$. In this case there are three important thresholds. Let us denote by $k$ the momentum of the hole, so the momentum of the particle is $k+q$. The lower threshold $\omega_L(q)$ of the particle-hole continuum is defined by a deep hole excitation composed of a particle with momentum  $k_F$ and a hole with $k=k_F-q$. The upper threshold $\omega_U(q)$ is defined by the particle-hole excitation which is symmetric about $\pi/2$, i.e.,  $k=(\pi-q)/2$. There is yet  a third  threshold $\omega_M(q)$, situated between $\omega_L(q)$ and $\omega_U(q)$, defined by a high-energy particle with momentum $k+q=k_F+q$  and a hole at $k_F$.

The exact $S(q,\omega)$ for the noninteracting model in the range $\pi-2k_F<q<2k_F$ is\cite{lieb}\be
S(q,\omega)=\frac{\theta(\omega_U(q)-\omega)[\theta(\omega-\omega_L(q))+\theta(\omega-\omega_M(q))]}{\sqrt{\omega^2_U(q)-\omega^2}}.\label{SVzero}
\ee
According to Eq. (\ref{SVzero}), $S(q,\omega)$ has  step discontinuities both at $\omega_L(q)$ and $\omega_M(q)$. The spectral weight jumps by a factor of two at $\omega_M(q)$
 because for $\omega>\omega_M(q)$ there are two choices of hole momentum $k$ corresponding to the same energy. For $\omega_L(q)<\omega<\omega_M(q)$, there is only one choice of $k$ for each $\omega$. At the upper threshold $\omega_U(q)$, $S(q,\omega)$ has a square root divergence which stems from the divergent joint density of states of a particle and a hole with the same velocity. 
 
 In the limit of half-filling, $k_F\to\pi/2$, the deep hole excitation and the high-energy particle excitation for the same $q$ become degenerate. The intermediate threshold merges with the lower threshold and one recovers a single step function at $\omega_L(q)$.\cite{muller2}

By analogy with the ansatz for the Heisenberg chain at zero magnetic field, M\"uller et al.\cite{muller1} proposed  that in the presence of repulsive interactions the two step discontinuities should become divergent edge singularities. This was argued to explain the double peak structure observed in the dynamical structure factor of spin chain compounds at finite field.\cite{heilmann} 

The picture of a line shape with two peaks at $\omega_L(q)$ and $\omega_M(q)$ is supported by the field theory approach to the edge singularities of $S(q,\omega)$. The divergence at the lower threshold is easily understood as an x-ray edge singularity for the deep hole excitation.\cite{pustilnik} For the upper threshold, it is known that, at half-filling, repulsive interactions turn the square-root divergence into a universal square-root cusp.\cite{pereira} The reason  is the resonant scattering between the high-energy particle and hole with the same velocity, which is   analogous to the problem of Wannier excitons in semiconductors.\cite{mahan}  Away from half-filling, the square-root divergence  still becomes a convergent singularity. However, the behavior near $\omega_U(q)$ is not exactly a square-root cusp because in the absence of particle-hole symmetry the high-energy particle-hole pair does not decouple from the Fermi surface modes.\cite{pereira} 
 
\begin{figure}
\includegraphics*[width=0.9\hsize,scale=1.0]{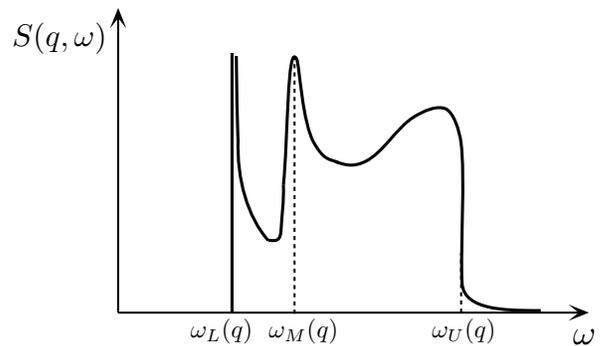}

\caption{Schematic illustration of the line shape of $S(q,\omega)$ for $\pi-2k_F<q<2k_F$ and small $V$. The most prominent features are the divergent singularity at the lower threshold and the rounded peak at the intermediate threshold.  \label{fig:lineshapeS}}
\end{figure}

The behavior near $\omega_M(q)$ is controlled by the x-ray edge singularity of a high energy particle excitation with $u<v$. Using the methods of Ref. \onlinecite{pustilnik}, it is easy to show that  $S(q,\omega)$ acquires a one-sided divergent power-law singularity above $\omega_M(q)$. The exponent of this singularity is different from the one at the lower threshold. Using the weak coupling expression for the phase shifts, we can show that the exponent at $\omega_M(q)$ to first order in $V$ is given by\be
\mu_M(q)=\frac{V(1-\cos q)}{\pi[\sin k_F-\sin(k_F+q)]}.
\ee
On the other hand, the exponent at $\omega_L(q)$ to  $O(V)$ is \be
\mu_L(q)=\frac{V(1-\cos q)}{\pi[\sin k_F-\sin(k_F-q)]}.
\ee
For $q>\pi-2k_F$, we have \be
\mu_M(q)>\mu_L(q).
\ee
However, based on the results of Sec. \ref{sec:DMRG}, we expect that below  half-filling the high-energy particle has a finite decay rate $1/\tau$ when $u<v$. Therefore, we predict that $S(q,\omega)$ has a rounded peak rather than a divergence at $\omega_M(q)$. A schematic picture of this line shape is shown in Fig. \ref{fig:lineshapeS}.

In the limit $k_F\to \pi/2$, the peak at $\omega_M(q)$ approaches the divergent singularity at $\omega_L(q)$. At the same time, $\mu_M$  becomes equal to $\mu_L$, as required by particle-hole symmetry. In addition, at half-filling the decay rate of the high-energy particle vanishes (see Sec. \ref{sec:decaylargek}). As pointed out in Ref. \onlinecite{cheianov}, the range of validity of the exponent $\mu_L$ shrinks to zero as $k_F\to\pi/2$. This was argued to lead to a discontinuity in the exponent, such that $\mu_L(k_F=\pi/2)\neq \mu_L(k_F\to\pi/2)$.  However, what happens is not  a crossover between two power-law singularities, as in the case of the spectral function,\cite{imamb2} but rather a collapse between two divergent singularities with the same exponent. Our scenario  predicts that $S(q,\omega)$ varies smoothly as $k_F\to\pi/2$. At $k_F=\pi/2$, one recovers the line shape with a momentum independent exponent at the lower threshold, $\mu_L=1-K$.\cite{pereira} Indeed, the results of Ref. \onlinecite{imamb3} show that the alternative exponent $\mu_L(k_F=\pi/2)$ proposed in Ref. \onlinecite{cheianov} is inconsistent with universal relations for the phase shifts in the low-energy limit and with the  SU(2) symmetry of the model for $V=2$. The correct exponent at half-filling is the one derived in Ref. \onlinecite{pereira}.

The line shape of $S(q,\omega)$ shown in Fig. \ref{fig:lineshapeS} can be tested by tDMRG calculations for the spin-spin correlation function of the XXZ model at finite magnetic field or by direct  computation of form factors from the BA, including negative-parity one-string excitations.\cite{maillet,pereiraJSTAT}
 
Here we have only discussed the nature of the thresholds defined by free particles and holes. We note that a recent BA work has shown that two-string excitations (two-particle bound states) give an important contribution  to the transverse dynamical structure factor of the spin-1/2 chain at finite magnetic field.\cite{kohno}

\section{Conclusion\label{sec:discuss}}

We have studied the spectral function of spinless fermions on a 1D lattice using  a combination of field theory methods, Bethe ansatz and time-dependent DMRG. We showed that x-ray edge type effective models are able to explain the long-time decay of the single-particle Green's function and the singularities of the spectral function calculated by tDMRG. 

The results can be summarized as follows. In the lattice model, the detailed line shape of the spectral function depends strongly  on momentum, density and interaction strength. Kinematically, the support of the spectral function has finite-energy lower thresholds for general momentum $k$ only if $k_F/\pi$ is a rational number. If $k_F/\pi$ is irrational, so that arbitrary Umklapp scattering processes  are incommensurate, the spectral function is nonzero at all energies. In any case, a general approximate picture for the line shape can be obtained by focusing on the possible divergent singularities on the single-particle energy or near multiparticle thresholds which involve bound states. 

Away from half-filling, the hole and particle contributions to the spectral function are remarkably different. For $k_F<\pi/2$, the hole spectral function exhibits a single divergent singularity above the single-hole energy. By contrast, the particle spectral function can exhibit one or two peaks. The single-particle peak is always present. In the regime where the velocity of the high-energy particle is larger than the Fermi velocity ($u>v$), the field theory predicts a divergence from both sides of the single-particle energy. For a generic model, this singularity is expected to be replaced by a broadened peak due to decay processes involving three-body collisions. For the integrable model we have found no compelling numerical evidence of broadening of the single-particle peak in this regime, even at fairly high energies, suggesting an exact power-law singularity. 

However, at higher values of momentum, for which the velocity of the high energy particle is smaller than the Fermi velocity, we found that the single-particle peak has a nonzero broadening. We interpret this in terms of a decay rate $1/\tau$ which is second order in the interaction strength. The broadening is expected to occur for integrable as well as non-integrable models for $u<v$. In real time, the corresponding particle Green's function decays exponentially. 

The second peak in the particle spectral function appears at high energies, above the single-particle energy, and is more pronounced for low densities and  momentum near $\pi$. This  peak has a  non-monotonic dispersion relation, with an energy minimum  at $k=\pi-k_F$. The nature of the second peak can be understood as the singularity at the lower threshold of the continuum defined by a two-particle antibound state and a free hole. 

The results of this work could be tested by measuring the spectral function of fully spin-polarized fermions in optical lattices. The most robust effect should be the observation of the bound state peak. In a photoemission experiment, where one probes the hole contribution to the spectral function, the bound state peak should be visible at high  densities (above half-filling) and for strong interactions. In this case, it can be interpreted as the process in which the outcoupled atom leaves behind a hole plus a particle-hole excitation. While the particle propagates freely, the two holes form a stable repulsively bound pair in the $p$-channel.

\acknowledgements
We thank J.-S. Caux for very useful discussions. This research was supported by NSERC (RGP and IA), CIfAR (IA), NSF under Grant No. PHY05-51164 (RGP) and NSF under grant DMR-0605444 (SRW).

\appendix
\section{Finite size spectrum\label{app:fs}}

In order to derive the finite size spectrum from the BA equations, we start by expanding the sum in Eq. (\ref{eq:BAeqfsfinal}) for large $L$ using the Euler-Maclaurin
formula\cite{korepin}\begin{eqnarray}
\rho\left(\lambda\right) & \approx & \frac{p_{0}^{\prime}\left(\lambda\right)}{2\pi}+\int_{B_{-}}^{B_{+}}\frac{d\lambda^\prime}{2\pi}\, K(\lambda-\lambda^\prime)\rho(\lambda^\prime)+\frac{\Phi_{p,h}\left(\lambda\right)}{2\pi L}\nonumber \\
 &  & +\frac{1}{24L^{2}}\left[\frac{K^{\prime}(\lambda-B_{+})}{\rho(B_{+})}-\frac{K^{\prime}(\lambda-B_{-})}{\rho(B_{-})}\right],\label{eq:Lieb_imp}\end{eqnarray}
where we introduced the Fermi boundaries $B_{\pm}=\lambda(I_{\pm})$.
We organize the solution to Eq. (\ref{eq:Lieb_imp}) by orders of $1/L$. We expand
$\rho(\lambda)$ up to  $O(1/L^2)$ in the form\begin{eqnarray}
\rho(\lambda) & = & \rho_{\infty}(\lambda|B_{+},B_{-})\pm\frac{1}{L}\rho_{imp}(\lambda,\lambda_{p,h}|B_{+},B_{-})\nonumber \\
 &  & +\frac{\rho_{1}(\lambda|B_{+},B_{-})}{24L^{2}\rho(B_{+})}-\frac{\rho_{1}(-\lambda|-B_{-},-B_{+})}{24L^{2}\rho(B_{-})},\label{eq:expandrhoBA}\end{eqnarray}
where the plus sign is for $\rho_{imp}(\lambda,\lambda_{p}|B_{+},B_{-})$, in the case of a particle, and the minus sign for $\rho_{imp}(\lambda,\lambda_{h}|B_{+},B_{-})$, in the case of a hole.
Substituting Eq. (\ref{eq:expandrhoBA}) into Eq. (\ref{eq:Lieb_imp}),
we find the integral equations for the terms in the expansion of $\rho(\lambda)$\begin{widetext}\begin{eqnarray}
\rho_{\infty}(\lambda|B_{+},B_{-}) & = & \frac{p_{0}^{\prime}\left(\lambda\right)}{2\pi}+\int_{B_{-}}^{B_{+}}\frac{d\lambda^\prime}{2\pi}\, K(\lambda-\lambda^\prime)\rho_{\infty}(\lambda^\prime|B_{+},B_{-}),\label{eq:lieb_inftyu}\\
\rho_{imp}(\lambda,\lambda^\prime|B_{+},B_{-}) & = & \frac{K(\lambda-\lambda^\prime)}{2\pi}+\int_{B_{-}}^{B_{+}}\frac{d\lambda^{\prime\prime}}{2\pi}\, K(\lambda-\lambda^{\prime\prime})\rho_{imp}(\lambda^{\prime\prime},\lambda^\prime|B_{+},B_{-}),\label{eq:inteqrhoimp}\\
\rho_{1}(\lambda|B_{+},B_{-}) & = & \frac{K^{\prime}(\lambda-B_{+})}{2\pi}+\int_{B_{-}}^{B_{+}}\frac{d\lambda^\prime}{2\pi}\, K(\lambda-\lambda^\prime)\rho_{1}(\lambda^\prime|B_{+},B_{-}).\label{eq:inteqrho1}\end{eqnarray}
\end{widetext}
Note that Eqs. (\ref{eq:lieb_inftyu}), (\ref{eq:inteqrhoimp}) and (\ref{eq:inteqrho1}) still contain terms of $O(1/L)$, implicit in the shifts of the Fermi boundaries due to the creation of the high-energy particle as well as the low-energy charge and current excitations. Let us denote by $B=\lambda(I_+=-I_-=N/2)$ the Fermi boundary for the ground state of $N$ fermions. Then, \begin{eqnarray}
\delta B_{+} & \equiv & B_{+}-B\sim O(1/L),\\
\delta B_{-} & \equiv & B_{-}+B\sim O(1/L).\end{eqnarray}
To zeroth order in $1/L$, Eq. (\ref{eq:lieb_inftyu}) becomes the Lieb equation Eq. (\ref{eq:dressingeq}).

We introduce the resolvent operator $\hat{L}$ formally by\be
\left(1+\hat{L}\right)\cdot\left(1-\frac{\hat{K}}{2\pi}\right)\left(\lambda,\lambda^\prime\right)=1,\ee
\be
L(\lambda|\lambda^\prime)-\frac{1}{2\pi}\int_{-B}^{+B}d\lambda^{\prime\prime}\, L\left(\lambda|\lambda^{\prime\prime}\right)K(\lambda^{\prime\prime}-\lambda^\prime)=\frac{K(\lambda-\lambda^\prime)}{2\pi}.\label{resolventL}\ee
To zeroth order in $1/L$, Eq. (\ref{eq:inteqrhoimp}) becomes Eq. (\ref{inteqforrhoimp}), with $\rho_{imp}(\lambda,\lambda^\prime)=\rho_{imp}(\lambda,\lambda^\prime|B,-B)$. By comparison with Eq. (\ref{resolventL}), we have\be
\rho_{imp}(\lambda,\lambda^\prime)  = L(\lambda|\lambda^\prime).\label{rhoimpp}
\ee

The ground state energy (without the impurity) in the thermodynamic limit is\begin{equation}
E_{GS}=L\int_{-B}^{+B}d\lambda\,\epsilon_{0}\left(\lambda\right)\rho_{GS}\left(\lambda\right).\end{equation}

Consider a state with a single high-energy particle $\lambda_p$ and general Fermi boundaries $B_\pm$. (The derivation is  similar for the case of a single hole.) The energy is\begin{equation}
E=\sum_{j}\epsilon_{0}\left(\lambda_{j}\right)+\epsilon_0(\lambda_{p}),\label{energy}\end{equation}
where the sum  runs over all quantum numbers $I_{j}$ between
$I_{-}$ and $I_{+}$. We expand Eq. (\ref{energy}) using the Euler-Maclaurin formula and obtain\bea
E&\approx&L\int_{B_{-}}^{B_{+}}d\lambda^\prime\,\epsilon_{0}\left(\lambda^\prime\right)\rho\left(\lambda^\prime\right)+\epsilon_0(\lambda_p)\nonumber\\
&&-\frac{1}{24L}\left[\frac{\epsilon_{0}^{\prime}\left(B_{+}\right)}{\rho\left(B_{+}\right)}-\frac{\epsilon_{0}^{\prime}\left(B_{-}\right)}{\rho\left(B_{-}\right)}\right].\label{energyEMcL}
\eea
The next step is to expand Eq. (\ref{energyEMcL}) up to $O(1/L)$ using the expansion of $\rho(\lambda)$ in Eq. (\ref{eq:expandrhoBA}) and assuming $\delta B_\pm \sim O(1/L)$. We find\bea
E&=&E_{GS}+\epsilon(\lambda_{p})-\frac{\pi v}{6L}\nonumber\\&&+\frac{\pi v}{L}\rho^2_{GS}(B)\left[ \left(L\delta B_{+}\right)^{2}+\left(L\delta B_{-}\right)^{2}\right].\label{eq:energyCFTshift}
\eea

The third term on the rhs of Eq. (\ref{eq:energyCFTshift}) is the standard
finite size correction to the ground state energy of a conformal field
theory with central charge $c=1$. The difference between the energy
of the excited state and the ground state energy  is\begin{equation}
\Delta E=\epsilon\left(\lambda_{p}\right)+\frac{\pi v}{L}\rho^2_{GS}(B)\left[ \left(L\delta B_{+}\right]^{2}+\left(L\delta B_{-}\right)^{2}\right] .\label{eq:almostthespectrum}\end{equation}

Now  we  evaluate the Fermi boundary shifts $\delta B_{\pm}$. From Eq. (\ref{eq:MIplusIminus}), we can write \be
\frac{N}{L}=n+\frac{\Delta N}{L}=\frac{I_{+}-I_{-}}{L}=\int_{B_{-}}^{B_{+}}d\lambda\,\rho\left(\lambda\right).\label{Nshifts}\ee
Similarly, from Eq. (\ref{eq:DIplusIminus}),
\bea
\frac{2D}{L}&=&\frac{I_{+}+I_{-}}{L}=\frac{1}{L}\left(\sum_{i<I_{-}}1\right)-\frac{1}{L}\left(\sum_{i>I_{+}}1\right)\nonumber\\&=&\int_{-\infty}^{B_{-}}d\lambda\,\rho\left(\lambda\right)-\int_{B_{+}}^{+\infty}d\lambda\,\rho\left(\lambda\right).\label{Dshifts}\eea
Expanding the rhs of Eqs. (\ref{Nshifts}) and (\ref{Dshifts}) to $O(1/L)$, we obtain\bea
\frac{\Delta N}{L}&=&\frac{n^p_{imp}}{L}+\left(\delta B_{+}-\delta B_{-}\right)\rho_{GS}(B)Z(B),\label{eq:diffdeltabs}\\
\frac{D}{L}&=&\frac{d^p_{imp}}{L}+\left(\delta B_{+}+\delta B_{-}\right)\rho_{GS}\left(B\right)\xi(B).\label{eq:sumdeltabs}
\eea
with $n^p_{imp}$ and $d^p_{imp}$ defined in Eqs. (\ref{nimpBA}) and (\ref{nimpBA}).
The function  $Z(\lambda)$ is the dressed charge defined by \begin{equation}
\left(1-\frac{\hat{K}}{2\pi}\right)\cdot Z(\lambda)=1.\end{equation}
The value of $Z(\lambda)$ at the Fermi boundary is related to the
Luttinger parameter by $K=Z^{2}(B)$.\cite{korepin} The function $\xi\left(\lambda\right)$ is defined by \begin{equation}
2\xi\left(\lambda\right)=1-\int_{B}^{\infty}d\lambda^\prime\, L\left(\lambda|\lambda^\prime\right)+\int_{-\infty}^{-B}d\lambda^\prime\, L\left(\lambda|\lambda^\prime\right).\label{csifunction}\end{equation}
One can  verify that the value of $\xi(\lambda)$ at the Fermi
boundary satisfies\begin{equation}
2Z(B)\xi(B)=1.\end{equation}
Using the definitions of $Z(\lambda)$ and $\xi(\lambda)$, together with Eq. (\ref{rhoimpp}), we can  write \bea
n^{p,h}_{imp}&=&\pm[Z(\lambda_{p,h})-1].\label{nZ1}\\
2d^{p,h}_{imp}&=&\pm[2\xi(\lambda_{p,h})-1].\label{dxi1}
\eea
Finally, substituting Eqs. (\ref{eq:diffdeltabs}) and (\ref{eq:sumdeltabs})
into Eq. (\ref{eq:almostthespectrum}), we arrive at the finite size spectrum\bea
\Delta E&=&\epsilon(\lambda_{p})+\frac{2\pi v}{L}\left[\frac{\Delta N-n^{p}_{imp}}{2Z(B)}\right]^{2}\nonumber\\&&+\frac{2\pi v}{L}Z^{2}(B)\left(D-d^{p}_{imp}\right)^{2}.\eea

\section{Phase shifts in terms of shift function\label{app:Ffunction}}
Here we prove the equivalence between our results\cite{pereira} in Eq. (\ref{nimpBA}) and (\ref{dimpBA}) and those of Cheianov and Pustilnik\cite{cheianov} in terms of the shift function. The shift function $F(\lambda|\lambda^\prime)$ is defined as the solution of the integral equation\cite{korepin}\be
\left(1-\frac{\hat{K}}{2\pi}\right)\cdot F(\lambda|\lambda^\prime)=\frac{\theta(\lambda-\lambda^\prime)}{2\pi}.\label{shiftF}
\ee
Upon application of the integral operator on Eq. (\ref{csifunction}), we find that $\xi(\lambda)$ obeys the equation\be
\left(1-\frac{\hat{K}}{2\pi}\right)\cdot \xi(\lambda)=\frac{1}{2}+\frac{\theta(\lambda+B)}{2\pi}, 
\ee
from which we obtain the relations\bea
\xi(\lambda)&=&\frac{Z(\lambda)}{2}+F(\lambda|-B),\\
\xi(-\lambda)&=&\frac{Z(\lambda)}{2}-F(\lambda|B).
\eea
Using Eq. (\ref{csifunction}), we can also show that\be
\xi(\lambda)+\xi(-\lambda)=1.
\ee
Hence from Eqs. (\ref{nZ1}) and (\ref{dxi1}), it follows that\bea
n^{h}_{imp}&=&F(\lambda|-B)-F(\lambda|B),\\
2d^{h}_{imp}&=&-F(\lambda|B)-F(\lambda|-B).
\eea
By taking the derivative of Eq. (\ref{shiftF}) with  respect to $\lambda$ and integrating either  inside or outside the Fermi boundaries we can show that \bea
Z(B)[F(B|\lambda)-F(-B|\lambda)]&=&Z(\lambda)-1,\\
2\xi(B)(B)[F(B|\lambda)+F(-B|\lambda)]&=&2\xi(\lambda)-1.
\eea
We then obtain\bea
n^{h}_{imp}&=&-Z(B)[F(B|\lambda)-F(-B|\lambda)],\\
d^{h}_{imp}&=&-\xi(B)[F(\lambda|B)+F(-B|\lambda)].
\eea
Finally, using $Z(B)=(2\xi(B))^{-1}=\sqrt{K}$, we can write the phase shifts in Eq. (\ref{gammasBA}) in the form\be
\gamma_{R,L}^h=\mp2\pi\sqrt{K} \,F(\pm B|\lambda_h).\label{c&p}
\ee
Eq. (\ref{c&p}) is equivalent to Eq. (14) of Ref. \onlinecite{cheianov} for any finite value of $B$ (away from half-filling).

\section{Two-spinon spectral weight for the zero field Heisenberg model near $q=\pi$}

In this appendix we determine what fraction of the total spectral weight is given by 
the two-spinon contribution to the dynamical structure factor in the zero field Heisenberg model of Eq. (\ref{heisenbergspinchain}) at $q\approx \pi$ and 
low frequencies.  The calculation is based on a combination of Luttinger liquid 
and exact integrability techniques. Band curvature effects are completely ignored in 
the Luttinger liquid treatment. 
This appears to be valid near $q=2k_F=\pi$ for the 
case of zero magnetic field (half-filled band in the fermion model; see Sec. \ref{sec:DSF}). 

Based on Luttinger liquid methods  it was argued\cite{AGSZ,Singh} 
that the asymptotic behavior of the equal time spin correlation function for 
the zero field Heisenberg model is
\be G(x,0)=\bra S^z_{j+x}S^z_j\ket\propto {(-1)^x\ln^{1/2}(|x|/a)\over |x|}.\label{Gx}\ee
Here $a$ is a number of order 1.
Later, the exact amplitude was determined
\cite{Affleck, Lukyanov}:
\be G(x,0)\to {(-1)^x\ln^{1/2}(|x|/a)\over (2\pi )^{3/2}|x|}.\label{Gx}\ee
 The field theory methods 
indicate\cite{AGSZ} that $G(x,t)$ is asymptotically a Lorentz scalar, depending 
only on $x^2-v^2t^2$. This is true including the logarithmic corrections 
since they arise from a Lorentz invariant marginal term in the low-energy effective 
Hamiltonian.  
Therefore, the extension to the time-correlation function is
\be G(x,t)\to {(-1)^{x}\ln^{1/2}[[x^2-v^2(t-i\alpha )^2]/a^2]\over 4\pi^{3/2}
[x^2-v^2(t-i\alpha )^2]^{1/2}}.\ee
Here  $\alpha >0$ is of order $1/J$. 
We now consider the Fourier transform, giving the structure function.
Let us first of all ignore the log factor, setting it to one.
Then we can do the integral exactly by changing variables to $vt\pm x$ and 
doing the two integrals separately, both of which are Gaussian. This gives:
\be S(q,\omega )\to \theta (\omega - v|q-\pi |) {1\over \sqrt{\pi}\sqrt{\omega^2-v^2(q-\pi  )^2}},\ee
for $q\approx \pi $ and small $\omega$.
We now include the log factor. 
\be S(q,\omega )\to \theta (\omega - v|q-\pi |){|\ln \{ a^2[\omega^2/v^2-(q-\pi )^2]\} |^{1/2}
\over \sqrt{\pi}\sqrt{\omega^2-v^2(q-\pi )^2}}.\label{Swl}\ee

To see this we can change variables to
\be (vt\pm x)(\omega /v\mp q) = u_{\pm},\ee 
giving:
\bea S(q,\omega )&=&{1\over 8\pi^{3/2}[\omega^2-v^2(q-\pi )^2]^{1/2}}\nonumber
\\
&&\times\int_{-\infty}^\infty du_+du_-
\frac{e^{i(u_++u_-)/2}}{[-(u_+-i\alpha )(u_--i\alpha )]^{1/2}}\nonumber \\
&&\times\left\{ \ln \left[\frac{-(u_+-i\alpha )(u_--i\alpha )}{a^2[(\omega /v)^2-(q-\pi )^2]}\right]\right\}^{1/2}.
\label{Swu}\eea
We now Taylor expand in powers of $\ln (-u_+u_-)$, integrating 
term by term. The leading term gives Eq. (\ref{Swl}). 

To further justify this expression and understand an important 
subtlety, we compare $S(\pi + p)$, the equal time structure factor, 
at small $|p|$, 
obtained either by Fourier transforming Eq. (\ref{Gx}) 
or by integrating Eq. (\ref{Swl}) over $\omega$.
\be S(q)=\int dx e^{-iqx}G(x)=\int {d\omega\over 2\pi } S(q,\omega ).\ee
The first approach gives:
\bea
S(\pi + p)&=&{1\over (2\pi )^{3/2}} \int_{|x|>a}dx{e^{-ipx}\ln^{1/2}(|x|/a)\over |x|}\nonumber\\&=&{2\over (2\pi )^{3/2}}\int_a^\infty 
dx \cos (px){\ln^{1/2}(x/a)\over x}.\label{x}\eea
 Note that this integral diverges at $|x|\to 0$ so it was necessary to 
introduce a cutoff on the integration range. We have chosen it to be $a$ 
for convenience but don't expect the leading behavior at $p\to 0$ to depend on this choice. 
We now approximate this integral by
\be
S(\pi + p)={2\over (2\pi )^{3/2}}\int_a^{c/|p|} 
dx {\ln^{1/2}(x/a)\over x},\label{x2}\ee
where $c$ is a constant of order 1. This is based on the approximation that, 
for small $p$, $\cos px$ is approximately constant and equal to 1 out to 
a large value of $x$ of order $1/|p|$. The integral can then be done 
exactly giving:
\be S(\pi +p)\to {4\over 3(2\pi )^{3/2}}\ln^{3/2}[c/(a|p|)]\label{S1}\ee
(This result was obtained in Ref. \onlinecite{Affleck}.)

Now  consider the second approach. 
Using $\omega_L\approx v|q-\pi |$, and introducing the high-energy cutoff, $D\propto 1/\alpha \propto J$, we obtain\bea S(\pi + p)&=&\sqrt{1\over \pi}
\int_{\omega_L}^{D} {d\omega\over 2\pi} {\ln^{1/2}[D^2/(\omega^2-\omega_L^2)]\over \sqrt{\omega^2-\omega_L^2}}\nonumber\\
&=&\sqrt{2\over \pi}
\int_0^{\sqrt{D^2-\omega_L^2}}{d\omega '\over 2\pi} {\ln^{1/2}(D/\omega ')\over \sqrt{(\omega ')^2+\omega_L^2}}\nonumber \\
&\approx& \sqrt{2\over \pi}
\int_0^{D}{d\omega '\over 2\pi} {\ln^{1/2}(D/\omega ')\over \sqrt{(\omega ')^2+\omega_L^2}}\nonumber\\
&\approx& \sqrt{2\over \pi}\int_{\omega_L}^D{d\omega '\over 2\pi \omega '}\ln^{1/2}(D/\omega ')\nonumber\\
&=&{4\over 3(2\pi )^{3/2}}\ln^{3/2}(D/\omega_L).
\label{S2}
\eea
Here we have changed variables to $\omega '\equiv \sqrt{\omega^2-\omega_L^2}$, 
in the first step, extended slightly 
the upper cutoff in the second and used the fact that the denominator 
can be approximated by $\omega '$ down to a value of order $\omega_L$ 
at which we may simply cut off the integral, in the third step.
  The fact that Eqs. (\ref{S1}) and 
(\ref{S2}) agree gives further confidence in these calculations. 

The factors of 2/3 that arise from the integrals in both approaches are rather 
subtle and easily missed by more crude approximations. In evaluating the 
integral of Eq. (\ref{x}), for example, it is tempting to follows the 
same procedure as we used in evaluating the integral in Eq. (\ref{Swu}), 
changing integration variables to $u\equiv xp$ and approximating 
$(\ln u - \ln ap )^{1/2}\approx \ln^{1/2}(1/ap)$. However, this 
would be incorrect, giving an answer larger by a factor of 3/2.  The 
fallacy with this approximation lies in the ultraviolet divergence of the 
integral at small $u$ of order $ap$. Since the integral receives important 
contributions from this range of $u$ it is not valid to drop the $\ln u$ term 
inside the square root. The reason why we could make this approximation in evaluating 
Eq. (\ref{Swu}) is that the integral is ultra-violet finite.

Now let us compare Eq. (\ref{Swl}) to the exact two-spinon result in Ref. \onlinecite{Karbach}. After correcting typos, precisely the same asymptotic behavior, 
Eq. (\ref{Swl}) is obtained, at $q\approx \pi$ and $\omega \to 0$ 
except that it is smaller by a factor of $\sqrt{C}$ 
where  $\sqrt{C}=0.849829$. Remarkably, the two-spinon approximation has 
the same combination of power-law and logarithmic singularities as does 
the exact result, but a smaller amplitude. 

Ref. \onlinecite{Karbach} attempts to derive a formula for $S(q)$ in Eq. (6.15). 
 However, 
 the tricky factor of 2/3 from the $\omega$ integral discussed above is missed. The correct two-spinon approximation to $S(q)$ is
\bea S(q)&\approx& {2\over 3}{m_0\over 2\pi}\left[-\ln (1-q/\pi )\right]^{3/2}\nonumber\\
&=&{2\over 3}\sqrt{2C\over \pi}{1\over 2\pi}\left[-\ln (1-q/\pi )\right]^{3/2},
\label{Sq2s}\eea
smaller by a factor of $\sqrt{C}\approx 0.85$ than the exact answer. The 
fact that the two-spinon approximation to $S(q)$ at $q\approx \pi$ underestimates the correct answer 
by exactly the same factor, 0.849829, as does the two-spinon approximation to $S(q,\omega )$ 
for $q\approx \pi$ and small $\omega$ is not surprising.  It merely indicates 
that the dominant contribution to the frequency integral is from low frequencies. 
As was discussed in Ref. \onlinecite{Karbach}, the total two-spinon intensity, 
integrated over all $q$ as well as all $\omega$ underestimates the 
exact answer (1/4) by a factor of 0.7289.  So, we see that the two-spinon approximation 
does somewhat better at $q$ near $\pi$ than at other values of $q$.

Finally, we remark that the asymptotic behavior of the finite size equal 
time correlation function, with periodic  boundary conditions, can be written \cite{Barzykin}:
\be G(x,0;L)\to {(-1)^x\ln^{1/2}[(L/\pi a)\sin (\pi |x|/L)]\over 
(2\pi )^{3/2}(L/\pi )\sin (\pi |x|/L)}.\ee
The $q=\pi$ finite size equal time structure function:
\be S(\pi ,L)\approx {2\over (2\pi )^{3/2}}\int_a^{L/2}dx 
 {\ln^{1/2}[(L/\pi a)\sin (\pi x/L)]\over 
(L/\pi )\sin (\pi x/L)},\ee
can be approximated as in Eq. (\ref{x2}), 
using the small $x$ approximation to the integrand with some large 
$x$ cut-off of $O(L)$:
\bea S(\pi ,L)&\approx& {2\over (2\pi )^{3/2}}\int_a^{cL}{dx\over x}\ln^{1/2}(x/a)\nonumber\\
&\approx& {4\over (2\pi )^{3/2}}\ln^{3/2}(L/a).\eea
Approximately this result was obtained\cite{Hallberg} by DMRG calculations, prior to the 
exact results, with $1/(2\pi )^{3/2}\approx 0.0632936 \ldots$ replaced 
by $0.06789$. We may obtain the two-spinon estimate of this quantity 
by replacing $|q-\pi |$ by $c/L$ in Eq. (\ref{Sq2s}). Again this is 
smaller than the exact result by the same factor of $0.85$ and thus 
smaller than the old DMRG result by 0.91 rather than being disturbingly larger 
than the DMRG result as was stated in Ref. \onlinecite{Karbach}, 
apparently due to missing the 2/3 factor.

\end{document}